\documentclass[%
preprint,
 amsmath,amssymb,
 aps,
]{revtex4-1}

\usepackage{graphicx}
\usepackage{dcolumn}
\usepackage{bm}
\usepackage{color}
\usepackage{amsmath}
\usepackage{comment}
\usepackage[utf8]{inputenc}
\usepackage{textcomp}
\usepackage{soul}
\usepackage{hyperref}
\usepackage{xr}
\usepackage{subfiles}
\usepackage{blindtext}

\newcommand{\changeshighlight}{black}  

\usepackage{float}
\newfloat{suppfig}{tbh}{losf}
\floatname{suppfig}{Supplementary Figure}

\begin{document}


\title{Light scattering control with neural networks in transmission and reflection}
\thanks{Supplementary videos available in: \href{https://www.osapublishing.org/oe/abstract.cfm?uri=oe-26-23-30911}{https://www.osapublishing.org/oe/abstract.cfm?uri=oe-26-23-30911}}

\author{Alex Turpin}
\thanks{These two authors contributed equally}

\author{Ivan Vishniakou}%
\thanks{These two authors contributed equally}

\author{Johannes D. Seelig}%
\email{Corresponsing author: johannes.seelig@caesar.de}
\affiliation{Center of Advanced European Studies and Research, 53175 Bonn, Germany}%

\date{\today}
             
\begin{abstract}
Scattering often limits the controlled delivery of light in applications such as biomedical imaging, optogenetics, optical trapping, and fiber-optic communication or imaging. Such scattering can be controlled by appropriately shaping the light wavefront entering the material.  
Here, we develop a machine-learning approach for light control. Using pairs of binary intensity patterns and intensity measurements we train neural networks (NNs) to provide the wavefront corrections necessary to shape the beam after the scatterer. 
Additionally, we demonstrate that NNs can be used to find a functional relationship between transmitted and reflected speckle patterns. Establishing the validity of this relationship, we focus and scan in transmission through opaque media using reflected light.
Our approach shows the versatility of NNs for light shaping, for efficiently and flexibly correcting for scattering, and
in particular the feasibility of transmission control based on reflected light.
\end{abstract}

\maketitle



\section*{Introduction}

When light propagates through a non-homogeneous and non-isotropic material its wavefront becomes distorted due to aberrations and scattering, resulting in an apparently random interference pattern of granular speckles \cite{2017:rmp:rotter,2017:natmeth:naji}. Such scattering conditions hamper the controlled delivery of light and the engineering of the PSF, which is a basic requirement for many applications \cite{2012:natcommun:dholakia,2010:natphoton:dholakia,2017:sa:yang,2012:prl:choi}.
To counteract this effect, methods based on shaping the light wavefront entering the scattering material have been developed. Wavefront shaping is typically achieved by using spatial light modulators (SLMs) \cite{2016:aop:forbes,2017:oe:cizmar} which, with their millions of degrees of freedom (pixels), allow focusing through diffusers \cite{2007:ol:vellekoop,2010:prl:gigan,2012:oe:piestun}, multimode fibers \cite{2011:oe:dileonardo,2012:natcommun:dholakia,2014:ol:choi}, and biological tissue \cite{2017:sciadv:yang,2015:scirep:park,2015:optica:yang,2015:natcommun:wang}. 
Different techniques have been developed to determine the appropriate wavefront corrections to be displayed on the SLM. 
\textcolor{\changeshighlight}{
The first demonstration of scattering control took advantage of iterative wavefront optimization \cite{2007:ol:vellekoop,2015:oe:vellekoop,2015:natphton:horstmeyer}, which approaches the targeted light distribution, typically a single or multiple focal spots, by updating the wavefront depending on the result after each optimization step\cite{2007:ol:vellekoop,2010:natphoton:dholakia,2012:natphoton:mosk,2012:PNAS:cui,2015:oe:vellekoop}. These feedback-based algorithms calculate the wavefront correction separately for each focal position or shape and can be optimized to achieve very fast focusing times \cite{2012:oe:piestun, 2017:ol:gigan}. 
}
A second approach, also typically used to control a single focus, is digital optical phase conjugation which uses interferometry to measure the scattered light field and reverses it with an SLM \cite{2010:oe:cui,2010:oe:psaltis,2012:natcommun:judkewitz,2013:scirep:yaqoob,2015:optica:yang}. This technique has the advantage of achieving update rates approaching the millisecond range needed for imaging in dynamic biological tissue \cite{2017:optica:wang}, while however requiring a focus or other guidestar to measure the appropriate correction. 
A third group of methods aims for describing and controlling the scattering process simultaneously across an entire field of view which was first achieved with the help of a transmission matrix
\cite{2010:prl:gigan,2010:natcommun:gigan,2017:optica:gigan}. For obtaining the transmission matrix, one needs to measure light phase, which, similar to digital optical phase conjugation, requires technically more demanding interferometric approaches. To simplify such experiments, computational methods for estimating incompletely measured information have been implemented which for example can infer the light phase from intensity measurements \cite{2015:oe:Dremeau,2017:ieee:metzler}.

Another set of computational techniques that, thanks to the development of programming frameworks together with the computational power of GPUs, is increasingly being applied in imaging and microscopy relies on machine learning (ML) \cite{2015:science:mitchell,2015:nature:hinton,2015:nature:waller,2018:prl:Deans} and in particular on NNs \cite{2015:optica:Kamilov,2017:optica_lensless:barbastathis,2017:optica:ozcan,2018:arxiv:shechtman, 2018:optica:psaltis}. The usefulness of these techniques \cite{2012:oe_ga:piestun,2014:jo:kner,2015:oe:kner,2017:jo:ding,2015:oe:tanida,2016:oe:tanida,2017:ao:tanida,2017:arxiv:barbastathis, 2017:arxiv:lyu, 1990:nature:sandler} has been demonstrated in the context of light scattering for image analysis \cite{2017:optica_lensless:barbastathis,2015:oe:tanida,2016:oe:tanida,2017:arxiv:barbastathis}, where the goal lies in the classification of an object across a scattering layer, or image reconstruction based on a predefined data set \cite{2017:arxiv:lyu}. 
\textcolor{\changeshighlight}{
In astronomy, NNs have been applied for the correction of weak scattering encountered when imaging through the atmosphere, for example for the control of multi mirror telescopes \cite{1990:nature:sandler}.
For light control, genetic algorithms, a class of iterative optimization algorithms, have been used for optimizing focusing across scattering materials \cite{2012:oe_ga:piestun,2014:jo:kner,2017:jo:ding}. 
}
Single- and multi-focus single-shot control (after training) over a $5 \times 5$ pixel area has been achieved using support vector regression  \cite{2017:ao:tanida}, but the reported small field of view, low signal to noise ratio, and long training times ($97\,\rm{min}$) are limiting for high-resolution PSF engineering.

While the methods outlined above allow focusing through scatterers in transmission, an additional set of challenges arises when the focal plane lies hidden behind or inside the scatterer, remote from direct optical access. For applications such as sensing, imaging or communication this is the more relevant configuration \cite{2018:prl:carminati}. For example in biological imaging, fluorescent signals or guidestars can be used to monitor excitation intensity inside or behind a scatterer (\cite{2012:PNAS:cui, 2015:scirep:park,2015:natphton:horstmeyer,2017:natmeth:naji}).
Particularly in the presence of strong scattering, however, these signals are often dim and generally have a limited photon budget \cite{2015:mc:betzig}. Alternatively, back-scattered excitation light can provide feedback about the beam \cite{2004:ol:denk,2012:oe:cui,2015:book:drexler}. These signals need to be additionally filtered to remove out-of-focus light and using various combinations of temporal, frequency, or spatial gating \cite{2015:book:drexler,2004:ol:denk,2012:oe:cui,2015:np:choi,2016:sa:aubry,2017:nc:choi,2018:optica:judkewitz} one aims for extracting photons that are scattered little and therefore retain image information. Since such weakly scattered photons disappear exponentially with depth, they in turn limit imaging depth.

However, even under strongly scattering conditions reflected (or backscattered) photons carry information about transmitted light \cite{2018:prl:carminati,2015:pra:carminati,2017:arxiv:bertolotti}. Mutual information between speckle patterns generated in these two opposite scattering directions indicates that reflected light might potentially be used to control transmitted light \cite{2018:prl:carminati,2015:pra:carminati, 2017:arxiv:bertolotti}. This would require that a functionally explicit relationship between these two scattering signals can be found and that the available information is sufficient for controlling one signal through the other.  So far, reflected light has been used to maximize the energy sent into a sample \cite{2013:prl:Yaqoob, 2015:oc:park}, but without control over the resulting light distribution, or required an embedded highly scattering target to achieve a localized light distribution \cite{2018:np:Choi}. Other schemes to take advantage of backscattered light have  been suggested in theoretical work \cite{2013:josa:rand, 2014:josa:rand}, but these concepts have so far not been implemented experimentally.

Here, we discuss how neural networks can be used to image through materials with different scattering characteristics such as glass diffusers, multi-mode fibers, or paper. First, we show that single-layer NNs (SLNNs) and multi-layer convolutional NNs (CNNs) can be trained to control the light distribution behind scattering materials with high accuracy. 
Second, we show that NNs can be used to find a functional relationship between transmitted and reflected light, i.e., they can predict transmitted speckle patterns from reflected speckle patterns with sufficient accuracy for light control through opaque materials. Taking advantage of this relationship we then show that NNs can be used for focusing in transmission using reflected light.

\section{Neural network approach for scattering control}

\begin{figure}[htbp]
\centering
\includegraphics[width=0.87\linewidth]{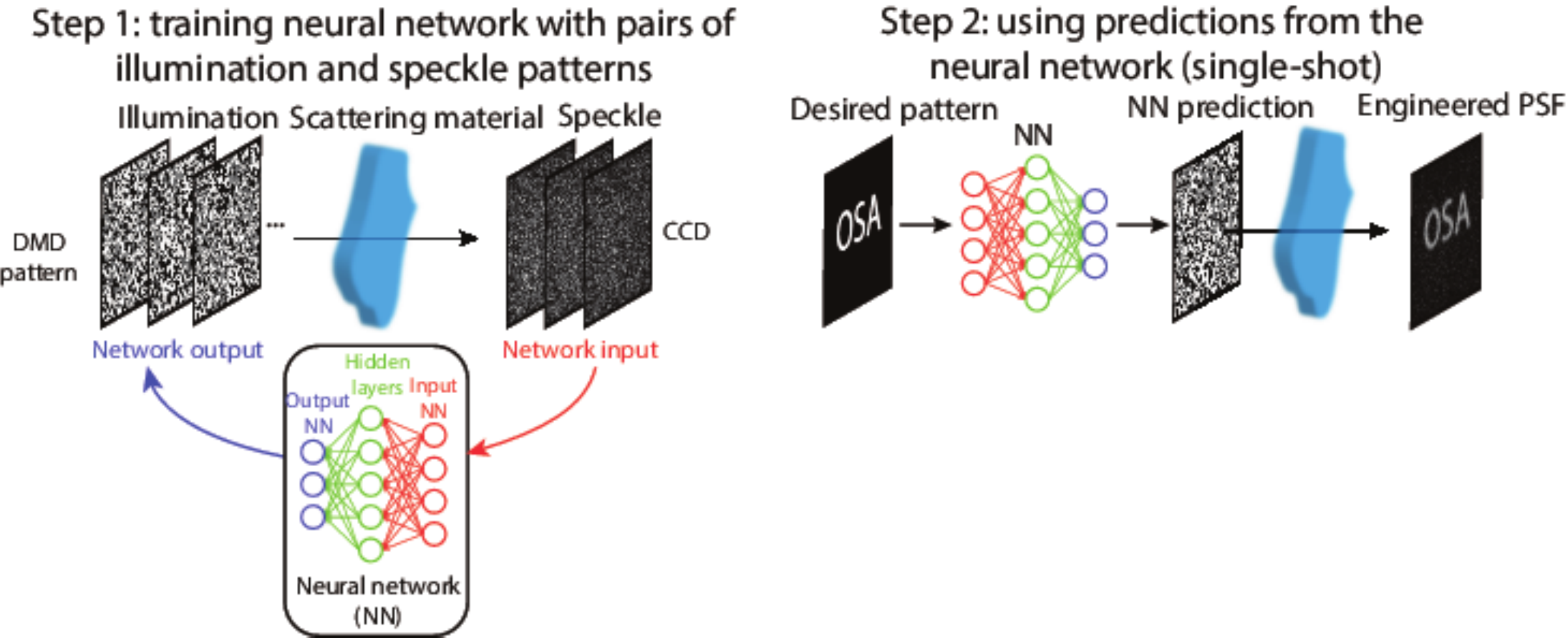}
\caption{Approach for light control through scattering media with NNs. A NN is trained with pairs of illumination and speckle patterns (illustration, see figures below for examples of actual illumination and speckle patterns), using the speckle patterns as input of the network and the illumination as output. Once the NN is trained, it is used to predict the illumination necessary to generate a target pattern after the scattering material. The predicted illumination is subsequently sent through the material resulting in the desired light pattern. 
}
\label{fig1a}
\end{figure}

We here first outline the underlying approach of using NNs for light control through a scatterer, which is also sketched in Fig.~\ref{fig1a}. In an initial step, we generate a dataset consisting of pairs of binary illumination patterns displayed on the SLM and corresponding speckle patterns recorded with a CCD camera after transmission through the scatterer \textcolor{\changeshighlight}{($64 \times 64$ macropixels for illumination patterns and $96 \times 96$ pixels for the CCD camera)}. 
These pairs of illumination and speckle patterns 
(typically on the order of 10000, but see below for training with a reduced number of patterns) 
are used to train the NNs as detailed below and in the Appendix, with the goal of inferring the relationship between the resulting scattered light distributions and the illumination patterns. 
We then feed the desired distribution into the trained NNs to predict the corresponding illumination pattern. This pattern is finally displayed on the SLM and the resulting light pattern is recorded with the camera.
Each pattern, $\mathbf{C\left( \mathbf{k} \right)}$, can be considered as a combination of plane waves with different wave vectors $\mathbf{k}$. This distribution of plane waves is modified by the scattering material through a function $F\left[\mathbf{C(\mathbf{k})}\right]$ in a deterministic way and results in the speckle pattern, $\mathbf{S}$, i.e. $ \mathbf{S} = F(\mathbf{C\left[\mathbf{k}\right]})$. Through training, the NN learns an approximation of the function $F$ needed to generate any light distribution after the scatterer. 

\begin{figure}[htbp]
\centering
\includegraphics[width=0.75\linewidth]{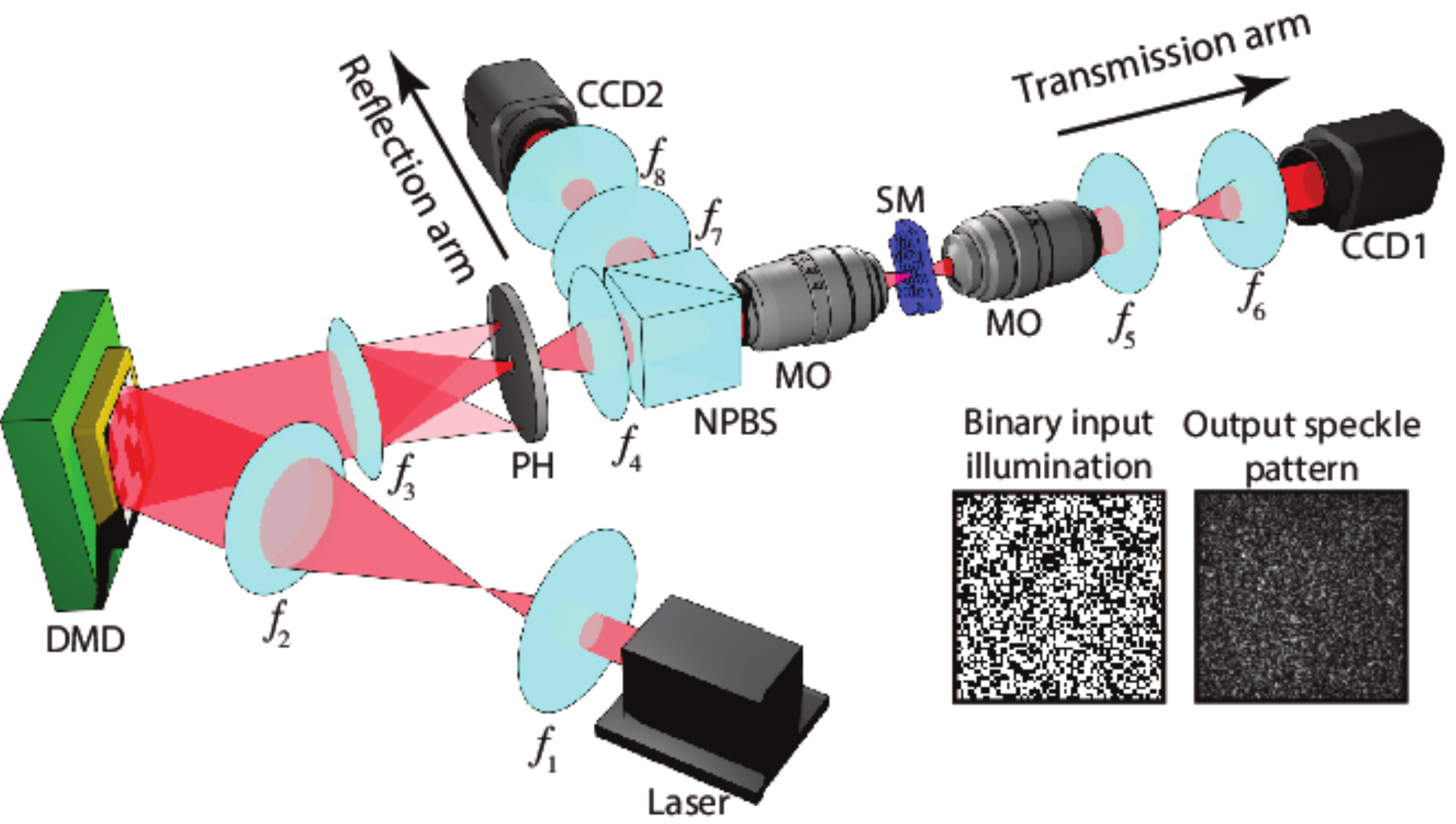}
\caption{Sketch of the experimental setup (see text for details). A DMD generates pseudo-random binary illumination patterns that are projected onto the back aperture of a microscope objective. After passing through the scattering material, light is projected onto a CCD camera by using a second identical microscope objective. A beam splitter placed before the first objective and a CCD camera are used to retrieve speckle patterns reflected by the sample for experiments with combined transmission and reflection. Bottom-right insets depict illustrations of illumination and speckle patterns. (PH: pinhole, NPBS: non-polarizing beam splitter, MO: microscope objective, SM: scattering material.)}
\label{fig1b}
\end{figure}


\section{Experimental set-up}

The experimental setup is schematically shown in Fig.~\ref{fig1b}. A laser beam ($\lambda=640\,\rm{nm}$, with an intensity of up to $P=100\,\rm{mW}$; iBeamSmart, Toptica) is expanded with a telescope ($f_1=15\,\rm{mm}$, $f_2=150\,\rm{mm}$) and sent to the SLM. (For the experiment shown in the Appendix we additionally included an optical isolator to minimize reflections into the laser (Thorlabs, IO-3D-633-VLP)). The SLM is a high-speed digital micromirror device (DMD, $768 \times 1024$ pixels, pixel size $=13.7\,\mu m^2$; model V-7000 from Vialux) allowing binary amplitude modulation at a maximum frame rate of $22.7\,\rm{kHz}$ and is used to display the illumination patterns (we have tested our system both with pseudo-random checkerboard-like patterns and with patterns obtained from Hadamard matrices) with typically $64 \times 64$ macropixels extending over the central $768 \times 768$ pixels of the DMD ($12 \times 12$ micromirrors per macropixel).
Two additional lenses ($f_3=200\,\rm{mm}$, $f_4=50\,\rm{mm}$) combined with a pinhole are used after the DMD to filter the maximum-intensity diffraction order mode and to demagnify and image the DMD onto the back aperture of the microscope objective (10X, $0.25\,\rm{NA}$, or 40X, $0.6\,\rm{NA}$ , WD=2.7-4.0 for one of the experiments shown in the Appendix, both from Olympus). The objective focuses the light beam through the scatterer (a glass diffuser, Thorlabs DG20-120, a step-index multimode fiber optic patch cable, Thorlabs M38L02; and a piece of white paper of 100 $\mu$m thickness) and a second identical microscope objective is used to collect the scattered light. Finally, a pair of lenses ($f_5=100\,\rm{mm}$, $f_6=75\,\rm{mm}$) in $2f$ configuration (or only a single lens with $f=60\,\rm{mm}$ for one of the experiments shown in the Appendix) images the back aperture of the second microscope objective (or the sample for one of the experiments shown in the Appendix) onto the CCD camera (acA640-750um, Basler), with a frame rate of $500\,\rm{fps}$ at full resolution of $480 \times 640$ pixels (pixel size $4.8\,\mu m^2$). Both microscope objectives and the scattering material are mounted on XYZ stages (omitted in Fig.~\ref{fig1b}) for aligning the system and moving the sample to different positions, as well as for displacing the image plane axially. In our experiments typically 10000 checkerboard patterns are uploaded to the internal memory of the DMD. Then, the projection of a pattern on the DMD triggers the frame capture of the CCD camera (transmission arm). The maximum frame rate of the DMD is $22.7\,\rm{kHz}$ and the maximum frame rate of the CCD camera is about $1000\,\rm{fps}$ at a resolution of $96 \times 96$ pixels, which allowed us to record the whole sequence in about $10\,\rm{s}$. 
We also note that our approach is valid for larger fields of view than those shown in the main figures ($20 \times 20\,\rm{\mu m^2}$, (see Appendix for details).
For experiments with reflected light, a non-polarizing beam splitter redirects the backscattered speckles towards a pair of lenses ($f_7=50\,\rm{mm}$, $f_8=25\,\rm{mm}$) in $2f$ configuration that image the back aperture of the first microscope objective onto a second CCD camera identical to and synchronized with the one used to capture the transmitted speckles (reflection arm). We used a computer wiht a Linux-Ubuntu operating system, an Intel Xeon CPU E5-1620 v4 @ $3.50\,\rm{GHz}$, 32Gb of DDR5 RAM memory, and a Nvidia Titan XP GPU possessing 3840 CUDA cores running at $1.60\,\rm{GHz}$ and with 12GB of GDDR5X memory running at over 11 Gbps.

\section{Neural networks for light control through scattering materials}

\begin{figure*}[htb]
\centering
\includegraphics[width=0.95\linewidth]{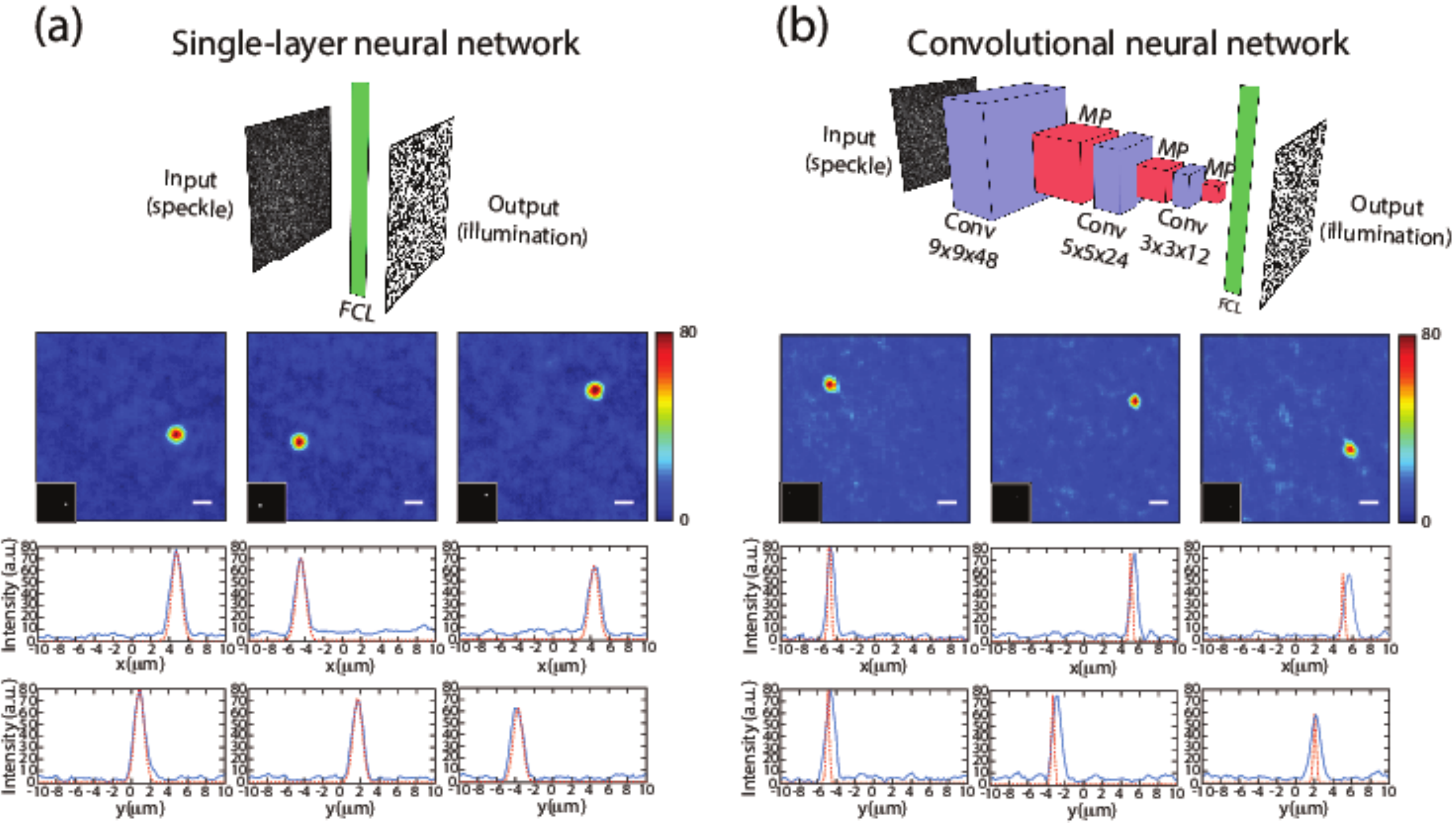}
\caption{
Focusing with neural networks. Top: Illustration of SLNN (a) or CNN (b) with speckle and illumination pattern. 
Bottom: Intensity distributions (first row) and intensity profiles through the foci along horizontal (second row) and vertical (third row) directions of Gaussian beams obtained at different positions after training (a) the SLNN and (b) the CNN. Red-dashed lines (\textcolor{red}{- -}) are the targeted intensity distributions that enter the NN, normalized to the corresponding experimental result. Scale bars = $2\,\rm{\mu m}$. FCL: fully-connected layer; Conv $n \times m \times p $: convolutional layer of $p$ kernels with dimensions $n \times m$; MP: max pooling operation reducing the previous element size. Color bars: intensity (a.u.).
}
\label{fig2a}
\end{figure*}

In Fig.~\ref{fig2a}(a) we demonstrate the ability of SLNNs to generate diffraction-limited Gaussian foci through a glass diffuser (as used for example in \cite{2015:natcommun:wang}) at different positions within the field of view. Top images schematically illustrate the NN architecture and training process, as detailed in the Appendix. Briefly, the SLNN connects all input to output channels through a single, so-called fully connected layer.  Below that, the first rows show the intensity distribution captured with the CCD camera, while the second and third rows display horizontal and vertical cross sections through the center of the focus. 
Insets and red-dashed lines show the position and shape of the target distribution that is fed into the neural network and for which the network then calculates the appropriate SLM mask (the target distribution is displayed normalized to the experimentally recorded intensity). 
The quality of the generated foci is analyzed with an automated procedure that generates spots at different positions placed in a grid throughout the whole field of view and measures the enhancement, defined as: 
$
\eta \equiv I_{\rm{focus}}/\langle I_{\rm{speckle}} \rangle,
$
where $I_{\rm{focus}}$ is the intensity at the generated foci and $\langle I_{\rm{speckle}} \rangle$ is the mean value of the background speckle \cite{2011:oe:mosk}. 

The images show an excellent agreement between the desired and recorded patterns (see also the Appendix for quantification of the match between the target and actual light distribution) with a signal-to-noise ratio $>10$ 
and an enhancement $\eta = 32 \pm 5$ for the 10X objective (see Appendix for scanning across the entire field of view) and an enhancement $\eta = 81 \pm 18$ for the 40X objective (see Appendix).
The time to achieve light control depends on the number of recorded frames and the training time. For the typical datasets of 10000 frames (\textcolor{\changeshighlight}{with a resolution of $64 \times 64$ macropixels on the DMD} and $96 \times 96$ pixels on the CCD, recorded at $1000\,\rm{Hz}$)
training on a single GPU required  $34\,\rm{s}$, and could be reduced down to $18\,\rm{s}$ while keeping an enhancement $\eta > 10$ (for the 10X objective see Appendix) .

While SLNNs are easy to implement and train, the underlying linearity limits their performance for many tasks \cite{2014:nn:schmidhuber}. A plethora of other network architectures have therefore been developed with the goal to improve over the performance of SLNNs. The most straightforward generalization of SLNNs combines multiple NN layers with connections between all neurons, resulting in a densely connected network. While such densely connected networks are not limited by linearity, the increased number of parameters also makes them more challenging to train, particularly for large data sets such as stacks of high resolution images. Network architectures  were therefore developed to take into account the structure of the underlying data and convolutional neural networks (CNNs) have emerged as one of the most successful solutions for image processing \cite{2014:nn:schmidhuber}. The typical architecture of a CNN consists of multiple convolutional layers that extract features across an entire field of view, interspersed with pooling layers that down-sample the image, and fully connected layers. While a large number of different networks are applied for different tasks, with a few to a few hundred convolutional layers \cite{2015:science:mitchell,2015:nature:hinton,2014:nn:schmidhuber}, we here found that a three-layer CNN (see Fig.~\ref{fig2a}(b) and Appendix for details) could be used for scattering control through a glass diffuser. To circumvent the difficulties of training nonlinear networks we pretrained the network with an autoencoder \cite{2014:nn:schmidhuber}, a network that compresses and then uncompresses the data into a close approximation of the input. (See also Appendix for focusing through paper with a different CNN architecture and training procedure.) The part of the network that was used for compression then served as the initial CNN for scattering control.

In Fig.~\ref{fig2a}(b) we demonstrate the ability of CNNs to generate diffraction-limited Gaussian foci through a glass diffuser (see Appendix for scanning across the field of view). The images again show a good agreement between the target pattern that was fed into the CNN (red dashed lines and inset) and the recorded patterns (see the Appendix for quantification) with a signal-to-noise ratio $>10$ and an enhancement $\eta = 3.6 \pm 0.9$ (measured over an ensemble of 25 different focus positions). For this particular application, CNNs reduced the number of network parameters by 80\% compared to SLNNs at the cost of lower enhancement with similar number of training samples. A larger enhancement of $\eta = 10 \pm 5$ was obtained with the 40X objective, as shown in the Appendix for focusing across the more strongly scattering paper. 

\subsection{SLNNs for point spread function engineering}
\begin{figure*}[h!]
\centering
\includegraphics[width=0.85\linewidth]{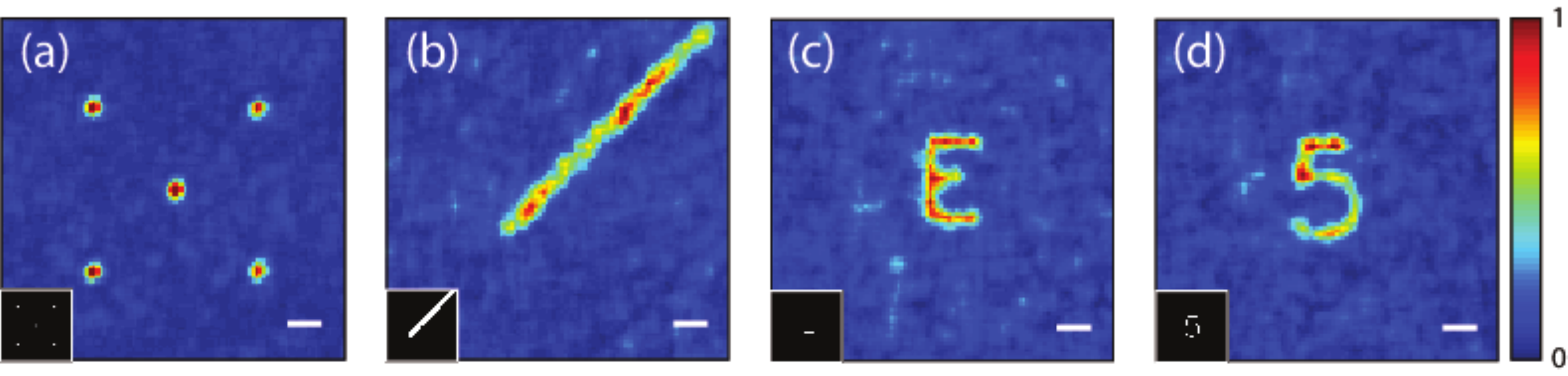}
\caption{Light control through a glass diffuser with a SLNN.  Normalized intensity patterns obtained after the glass diffuser with the SLNN: (a) five Gaussian foci; (b) a line at $45^{\circ}$; (c) the letter "E"; and (d) the number "5"; Insets show the desired light distribution. Scale bars = $2\,\rm{\mu m}$. Color bar: intensity (a.u.) normalized for each image.}
\label{fig3}
\end{figure*}

Since the SLNN is a linear network, we reasoned that after training it should be able to take advantage of the linearity of light scattering in non-absorbing media to generate arbitrary light distributions.
To demonstrate the validity of our approach for controlling the light intensity distribution after the scatter we generated in Fig.~\ref{fig3} a variety of non-trivial shapes using SLNNs. Again, there is an excellent correspondence between the target distribution that enters that network (insets) and the recorded patterns. 
We note that thanks to the high frame rate of the DMD ($22.7\,\rm{kHz}$), alternatively, any shape can be generated with high fidelity by painting it spot by spot, e. g. similar to approaches for trapping ultra cold atoms \cite{2009:njp:boshier} or optogenetics \cite{2009:pnas:tank}, as shown in \textcolor{blue}{Visualization 1}, \textcolor{blue}{Visualization 2}, \textcolor{blue}{Visualization 3}, \textcolor{blue}{Visualization 4}, \textcolor{blue}{Visualization 5}, \textcolor{blue}{Visualization 6}.

\subsection{SLNNs for light control through optical fibers}
\begin{figure*}[h!]
\centering
\includegraphics[width=0.85\linewidth]{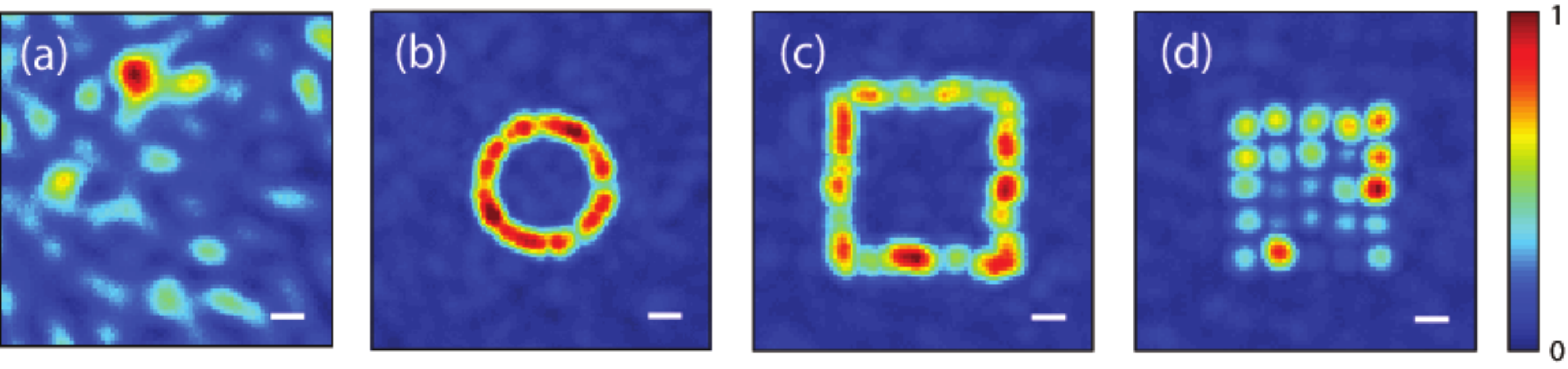}
\caption{Neural networks focus light through multimode fibers. Normalized transverse maximum intensity projection of the light field (see \textcolor{blue}{Visualization 7}, \textcolor{blue}{Visualization 8}, \textcolor{blue}{Visualization 9} for further information) after a multimode fiber when (a) no correction is applied and when a single focus is scanned (maximum intensity projection) along (b) a circle, (c) a square, and (d) an array of $5 \times 5$ points. Scale bars = $2.3\,\rm{\mu m}$. Color bar: intensity (a.u.) normalized for each image.}
\label{fig4}
\end{figure*}

Our system is suited well to correct for scattering in materials with slow dynamics (on the order of a few tens of seconds, see Appendix),
such as optical fibers \cite{2011:oe:dileonardo,2017:oe:piestun,2017:arxiv:piestun, 2018:oe:Zhao}. In particular, multimode optical fibers are ideal for applications in imaging and optogenetics, but modal dispersion and cross-talk distribute light into an apparently random speckle pattern. We therefore tested the performance of SLNNs for controlled light delivery through multimode fibers. In Fig.~\ref{fig4} and \textcolor{blue}{Visualization 7}, \textcolor{blue}{Visualization 8}, \textcolor{blue}{Visualization 9} a single focus is scanned \textcolor{\changeshighlight}{($\eta = 10 \pm 3$)} with different paths across the field of view of the fiber (including a circle, a square, and a $5 \times 5$ array of points), demonstrating that SLNNs are able to precisely control light through optical fibers. 

\section{Neural networks find functional relationships between transmitted an reflected speckle patterns}
While most methods for focusing light through strongly scattering media rely on measuring transmitted light (as in the experiments described so far), many applications could benefit from using reflected light. Towards that goal we tested whether neural networks (NNs) can take advantage of mutual information between transmission and reflection images \cite{2018:prl:carminati,2015:pra:carminati, 2017:arxiv:bertolotti} for light control. In the following we show that with the help of NNs it is indeed possible to find a functional relationship between reflected and transmitted speckle patterns to control transmitted light using reflected light. 

\begin{figure*}[h!]
\centering
\includegraphics[width=0.85\linewidth]{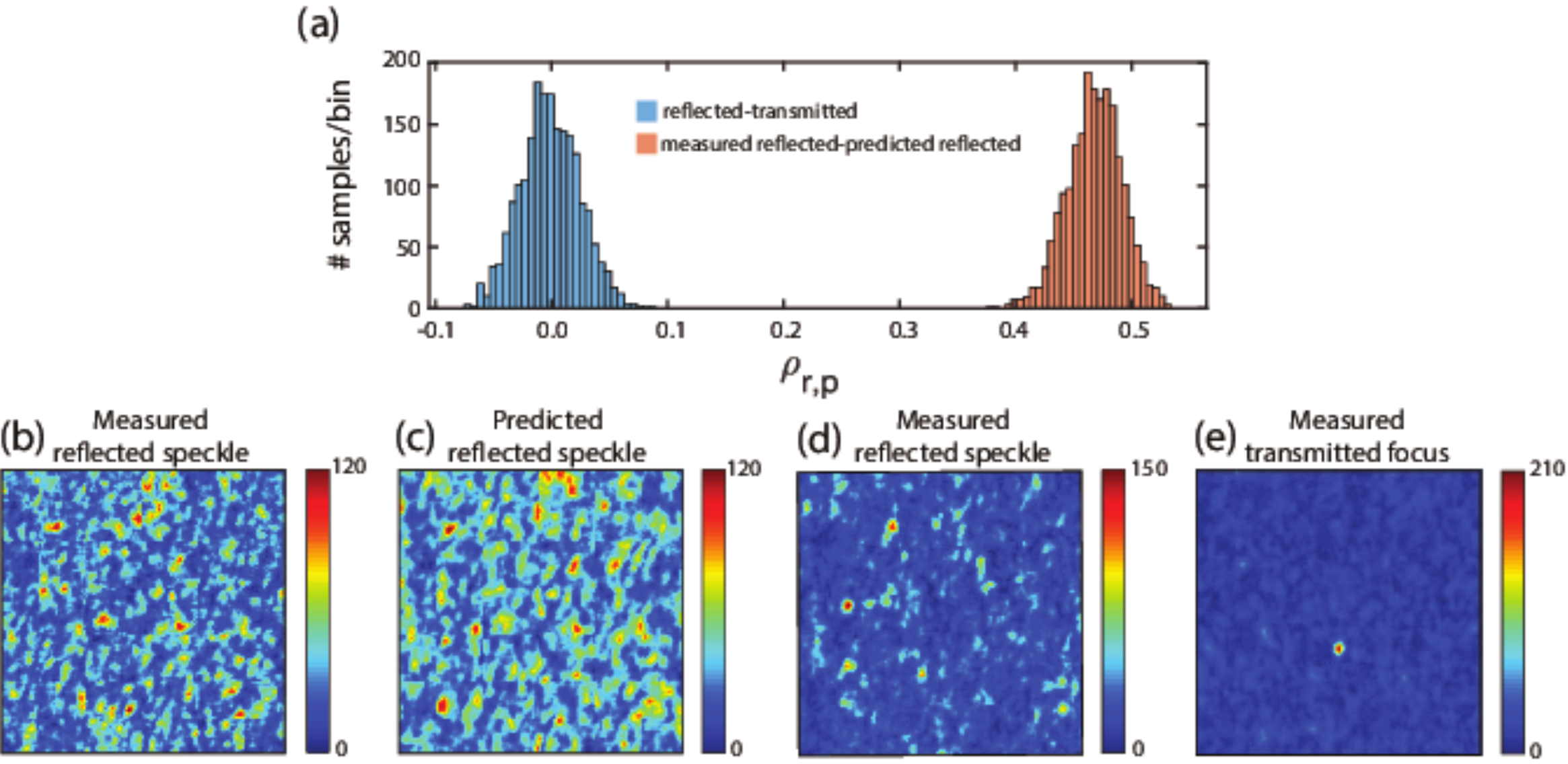}
\caption{Neural networks find functional relationships between transmitted and reflected speckle patterns. (a) Histogram of the Pearson correlation coefficient $\rho_{r,p}$ between the measured reflected and transmitted speckle patterns (in blue) and the measured and predicted reflected speckle patterns (orange). Number of bins: 30. Total number of pairs of samples used: 2000. 
Bottom: Examples of the normalized transverse intensity distributions of the light field of the (b) measured reflected speckle pattern, (c) predicted reflected speckle pattern, and (d) measured reflected speckle pattern while focusing in transmission (e).}
\label{fig5}
\end{figure*}

For this experiment we simultaneously recorded transmitted and reflected light by adding a non-polarizing beam splitter, a pair of imaging lenses and a CCD camera to the setup, as shown in Fig.~\ref{fig1b}. To achieve good signal to noise ratio of transmitted as well as reflected speckle patterns, we used paper as scattering material, which was more strongly scattering than the glass diffuser \cite{2016:sa:aubry} and led to an increased amount of backscattered light. Sets of simultaneously recorded transmitted and reflected speckle patterns (with size of $128  \times 128$ pixels) were then generated by illuminating the sample with a series of checkerboard projections ($64  \times 64$). Once the speckle patterns were recorded, we trained a SLNN (SLNN1) to find the relationship between transmitted and reflected light. 
To quantify the performance of this network we used the Pearson correlation coefficient as a similarity measure \cite{2012:book:goshtasby} between transmission speckle patterns predicted by the network and measured transmission speckle patterns. Fig.~\ref{fig5}(a) shows the histogram of these correlation coefficients and the correlation coefficient between transmitted and reflected speckle patterns for comparison. Figs.~\ref{fig5}(b) and (c) show, respectively, an example of measured and predicted speckle patterns with median correlation ($\rho_{} = 0.50$), while Figs.~\ref{fig5}(d) and (e) show an example of measured reflected speckle pattern when a corresponding focus is generated in transmission. Note that when focusing in transmission the intensity of the speckle pattern and the number of speckle grains in reflection decrease.

\begin{figure*}[h!]
\centering
\includegraphics[width=0.75\linewidth]{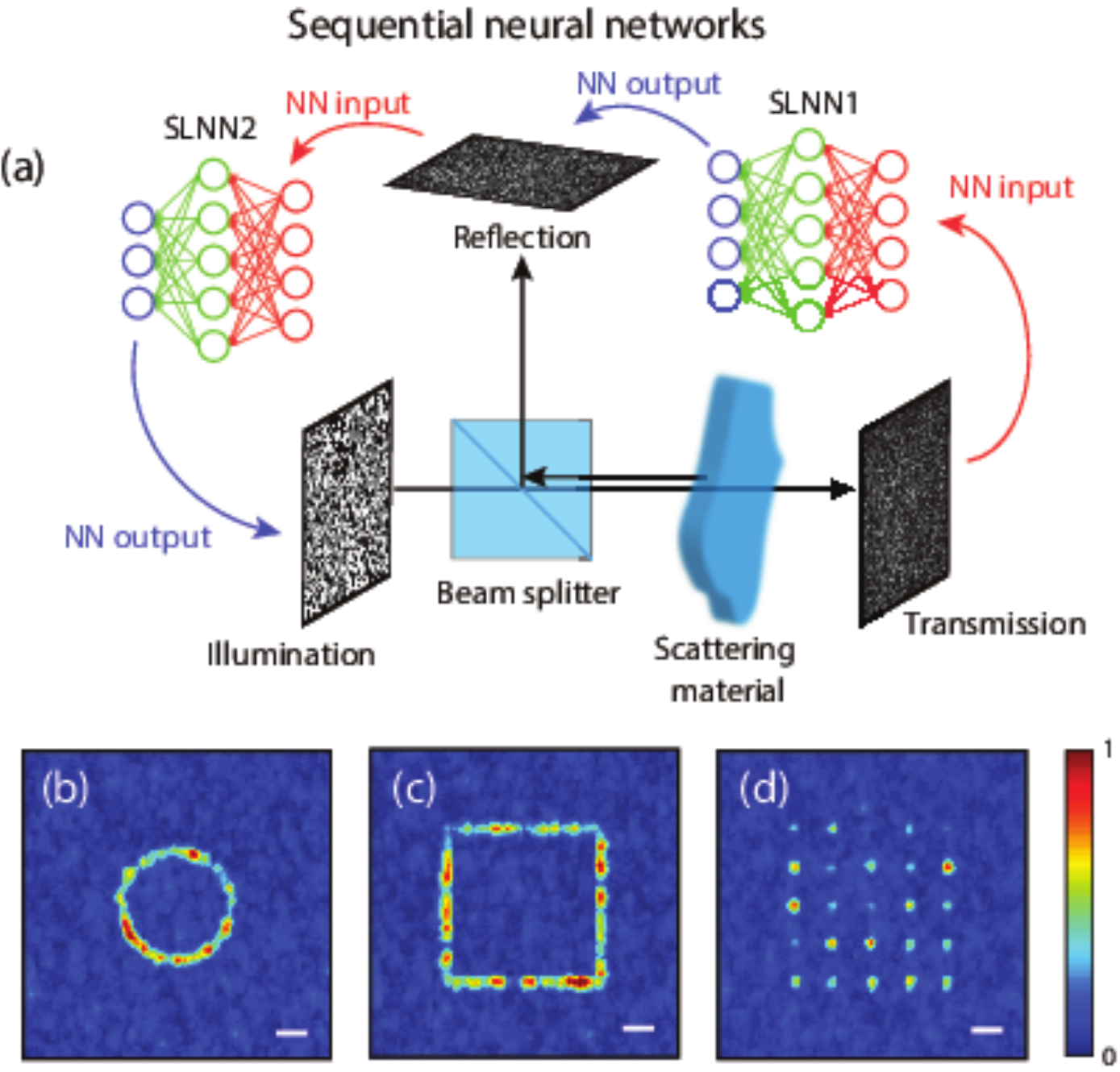}
\caption{Focusing and scanning in transmission using reflected light. Illustration of network approach to control transmitted light using reflected light. A SLNN is trained to learn the relationship between simultaneously recorded transmitted and reflected speckle patterns. After training of the network, it is sufficient to train a second SLNN to relate reflection to illumination for controlling light through the combined network. Bottom: Normalized transverse maximum intensity projection of the light field after a sheet of paper when a single focus is scanned (maximum intensity projection) along (a) a circle, (b) a square, and (c) an array of $5 \times 5$ points. Scale bars = $2.7\,\rm{\mu m}$. Color bar: intensity (a.u.) normalized for each image.}
\label{fig6}
\end{figure*}

To take advantage of this relationship between transmitted and reflected light for light control, we trained a second independent network (SLNN2 in Fig.~\ref{fig6}(a)) to infer the relation between reflected speckles and illumination patterns, similar to the training of the SLNN in the transmission configuration in the previous sections of the article (that is, with the reflected speckles as input of the SLNN and the illumination patterns as output, see Appendix for details). 
Combining these SLNNs, as shown in Fig.~\ref{fig6}(a), allowed us to form transmission foci by only taking advantage of reflected light, based on SLNN1 relating reflected to transmitted speckle patterns. In Figs.~\ref{fig6}(b)-(d) we show, respectively, that we can scan a circle, a square, and a grid, demonstrating full control of transmitted modes using reflected modes over the entire field of view. This additionally demonstrates that the predicted speckle patterns (Fig.~\ref{fig5}) are sufficiently accurate for high-resolution light control. The measured enhancement in this case was $\eta = 12 \pm 4$ (measured over an ensemble of 25 different focus positions). Note also that even though paper is more strongly scattering
\cite{2016:sa:aubry} than the glass diffuser, 
this does not hinder the SLNN from light control. Focusing through paper with a SLNN as reported in the previous sections is shown in the Appendix and \textcolor{blue}{Visualization 10}, \textcolor{blue}{Visualization 11}, \textcolor{blue}{Visualization 12}.

\section{Discussion and conclusions}

In summary, we showed that NNs can be used to efficiently shape light through a variety of media with different scattering characteristics (Figs.~\ref{fig2a} -- ~\ref{fig4}, 8--11). Once the NNs are trained, we achieve real-time, single-shot light control through the scattering material with high fidelity, in a fashion similar to transmission matrix approaches \cite{2010:prl:gigan,2015:spie:loterie,2017:optica:gigan}. Specifically, we demonstrated the ability of SLNNs to focus and scan light through glass diffusers, multimode fibers, and paper, and to generate arbitrary light distributions through glass diffusers.  We further showed that nonlinear networks, specifically CNNs, can focus light through a glass diffuser and paper. 

In a second set of experiments, we demonstrated that with the help of two networks, one establishing an explicit functional relationship between light that is transmitted through a scatterer and light that is reflected, and one relating reflected light to illumination patterns, we can control transmission using reflection at diffraction limited resolution. SLNNs  therefore prove to be well suited to take advantage of a recently described mutual information between transmitted and backscattered light for light control \cite{2018:prl:carminati,2015:pra:carminati, 2017:arxiv:bertolotti}. 

To compare the performance of the NN method for focusing in transmission with other schemes, we quantified the enhancement as in \cite{2011:oe:mosk} and obtained values similar to those reported for intensity-only modulation for the SLNN (see Fig. 9 in the Appendix for scanning through paper with the SLNN) \cite{2011:oe:mosk,2014:jo:kner, 2015:oe:Dremeau} and lower values for the CNNs (but still with a sufficient SNR and enhancement for imaging applications, see also Fig. 11 in the Appendix for an example of scanning through paper with CNNs), with the caveat that a direct quantitative comparison needs to take into account the specific combination of scatterer, optical setup (we used lower N.A. objectives than many of the reports with higher enhancement, and as expected also measured lower enhancement with the 10X objectives as compared to the 40X objectives), and the number of controlled modes. The maximum number of controlled modes in our experiments was 4096 ($64 \times 64$). While the SLM has $768 \times 1024$ pixels, the number of controllable modes depends on the memory of the GPU of 12 GB which needs to accommodate the NN model and a single batch of training data, in our case typically 150 frames. This limit could however be overcome by using multiple GPUs. 

To increase the enhancement and light control one could in addition modulate phase (see p. 42 of \cite{2017:rmp:rotter} for the effect on enhancement and signal-to-noise ratio) and the NN approach could be extended to any combination of stimulus-response pairs, including phase or polarization on either the detection or projection side, or both. An advantage of only using binary intensity modulation and intensity measurements \cite{2015:oe:Dremeau}, is that it simplifies the setup compared to approaches that also rely on phase information. Additionally, although we used monochromatic light, our approach could also be used with pulsed light \cite{2018:lsa:booth}. 

For applications, the time it takes to achieve light control is critical. For the transmission matrix approach as well as the NN approach this time can be broken down into two parts, the time for acquiring the data and the time to compute the wavefront correction. The acquisition time is ultimately limited by the number of required frames. Typical numbers for the transmission matrix approach in recent reports range from 4000 \cite{2017:optica:gigan} to 12000 \cite{2017:arxiv:piestun} which is similar to the number of frames used in our experiments, which was typically 10000 or less, see Figs.~12 and 13 (see also \cite{2018:oe:Zhao} for measuring the transmission matrix using 5000 sample pairs). The time-limiting factor in our experiments was training of the NNs. For the largest data sets the time required for training was less than 35 seconds for the SLNNs and less than 50 seconds for fine-tuning the CNNs or 150 seconds for training the second CNN architecture (Fig. 11).
To accelerate the process we tested training with a reduced amount of data (Fig.~13) which sped up training at the cost of lower enhancement. The shortest training time on a single GPU with the SLNN that lead to a focus with significant enhancement ($\eta > 10$ for the 10X objectives) was obtained with 5000 frames in 18 seconds. 
For comparison, the time required for calculating the transmission matrix varies for different techniques, from a simple Hermitian conjugation operation, to computationally more demanding approaches which require 15 seconds matrix multiplication on a GPU \cite{2017:arxiv:piestun}.
While some methods that optimize a single mode can be very fast (for example $33.8\,\rm{ms}$ in \cite{2012:oe:piestun}), this still results in a comparable correction time for a full field of view (of about 5 minutes for $96 \times 96$ pixels in this example).

Further improvements of the NN approach could be achieved by optimizing the network architectures. Here, we compared two basic networks, SLNNs and CNNs. 
SLNNs take advantage of the linearity of scattering (as does the transmission matrix approach) and therefore can generalize from speckle patterns to arbitrary light distributions.  Multi-layer NNs in contrast need to be specifically designed and trained to generate a desired type of light distribution. That CNNs are not constrained to the linearity of the underlying scattering process also might explain their worse performance in our experiments which could potentially be remedied with a larger training data set. At the same time, that multi-layer NNs are independent of assumptions about the underlying physical model (such as linearity of scattering) and can efficiently reduce the dimensionality of the images through convolutional layers as well as lower the number of parameters required for training (by 80\% compared to the SLNNs in our case), will likely prove advantageous for applications for example in nonlinear situations \cite{2017:optica:silberberg}. 

For many applications of light control through scattering media, such as imaging, sensing or communication, it will be necessary to develop methods that can work with reflected light \cite{2018:prl:carminati}. For example in biological microscopy, fluorescence signals can serve as feedback for scattering correction \cite{2015:natphton:horstmeyer,2017:natmeth:naji}, but they require labeling of the sample and are often dim, particularly before wavefront correction. Other schemes for light control in tissue resort to the assistance of acoustics waves \cite{2011:np:wang,2017:sa:yang} but do not achieve diffraction limited optical resolution \cite{2015:natphton:horstmeyer}. The most broadly applicable implementation for wavefront correction takes advantage of backscattered light as for example in optical coherence tomography or related approaches \cite{2015:book:drexler,2004:ol:denk,2012:oe:cui,2015:np:choi,2016:sa:aubry,2017:nc:choi,2018:optica:judkewitz, 2018:ol:Kuschmierz}. However, ultimately the availability of weakly scattered photons is limiting the imaging depth of these methods and ways to take advantage of strongly scattered light are therefore needed. Strategies for light control using strongly scattered, reflected light have indeed been developed  \cite{2013:prl:Yaqoob, 2015:oc:park, 2018:np:Choi} for maximizing the energy delivered into the material \cite{2013:prl:Yaqoob, 2015:oc:park} or an embedded strongly scattering target \cite{2018:np:Choi} without, however, exerting full independent control over the transmitted modes. 
We here took advantage of mutual information between transmitted and reflected speckle patterns \cite{2018:prl:carminati,2015:pra:carminati, 2017:arxiv:bertolotti}
and used NNs to show that it is indeed possible to control transmitted light with reflected light with sufficient accuracy for high-resolution focusing and scanning (Fig.~\ref{fig6}). We achieved this by using NNs to establish an explicit functional relation  between transmitted and reflected speckle patterns (Fig.~\ref{fig5}). That such a relationship can be established (with a  linear network) could not necessarily be expected based on the mutual information relationship in \cite{2018:prl:carminati}.
 
The limitation of the current approach for applications is that it first requires characterizing the transmission and reflection properties of the scatterer for the specific field of view, which still requires unobstructed access to the focal plane behind the scatterer. How could this limitation be overcome? One of the distinctions of neural networks is their ability to generalize. A potential avenue would therefore be to train appropriate NN models on sufficiently broad training sets and to adapt these models to the specific sample or field of view \cite{2018:arxiv:tian}, e.g. using backscattered light. For example, CNNs are the building blocks for many of the more advanced network techniques that analyze novel visual scenes based on previously learned data sets \cite{2015:science:mitchell,2015:nature:hinton,2014:nn:schmidhuber}, and such methods might also be harnessed for light scattering. 
 
Independent of this, the simplicity, effectiveness and flexibility of the method presented here makes it suitable for scattering control in transmission or in reflection as well as for the further analysis of the relationship between transmission and reflection in scattering materials. 

\section*{Appendix}

\begin{figure*}[htb]
\centering
\includegraphics[width=0.75\linewidth]{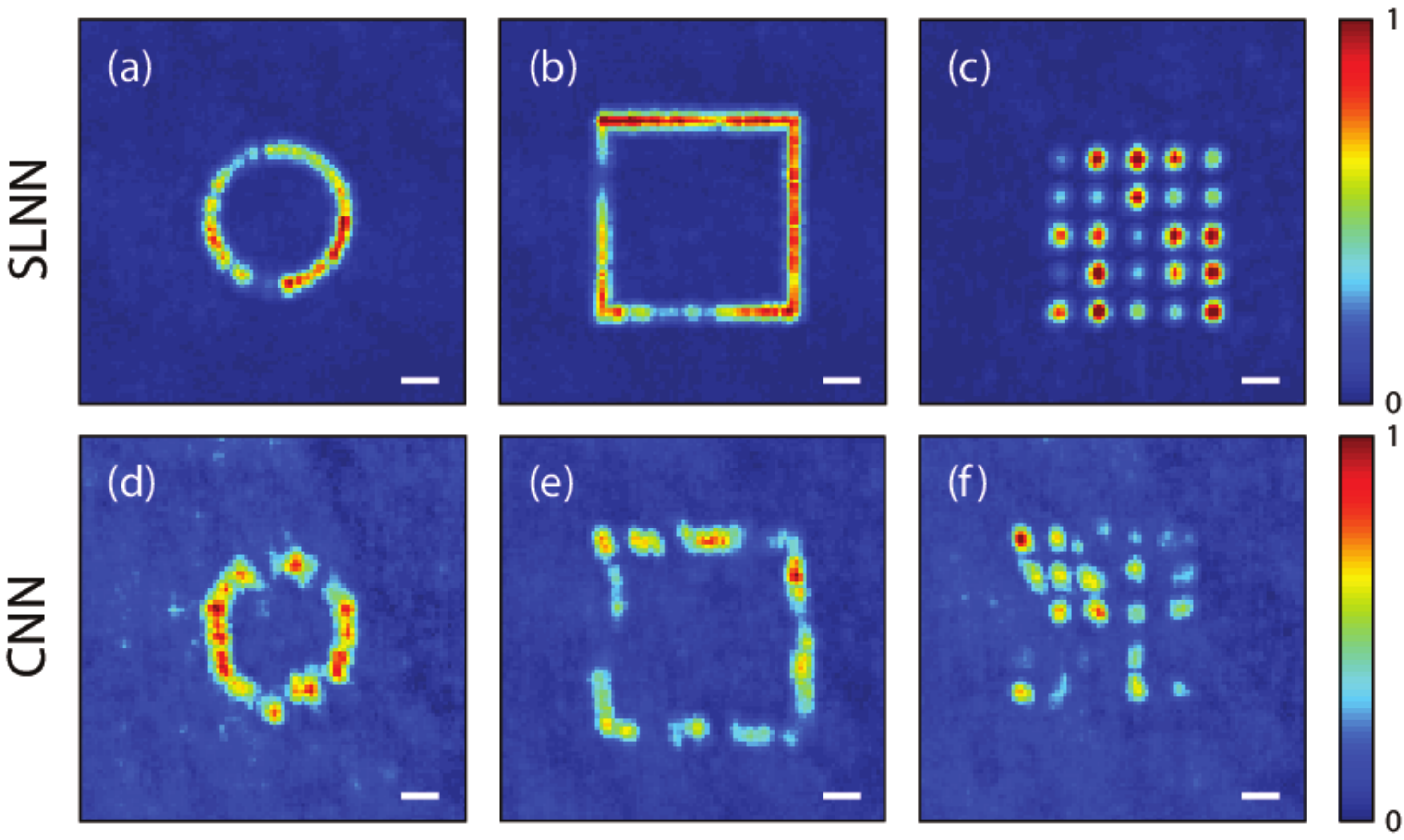}
\caption{Single-focus scanning allows time-averaged pattern projection. Patterns obtained when a single focus is scanned following (a/d) a circle (128/96 scanning points), (b/e) a square (256/256 scanning points), and (c/f) a grid of $5 \times 5$ points with the SLNN (first row) and the CNN (second row). Color bars: intensity (a.u.) normalized for each image. Scale bars:$2.3\,\rm{\mu m}$}
\label{sfig1}
\end{figure*}

\begin{figure*}[htb]
\centering
\includegraphics[width=0.75\linewidth]{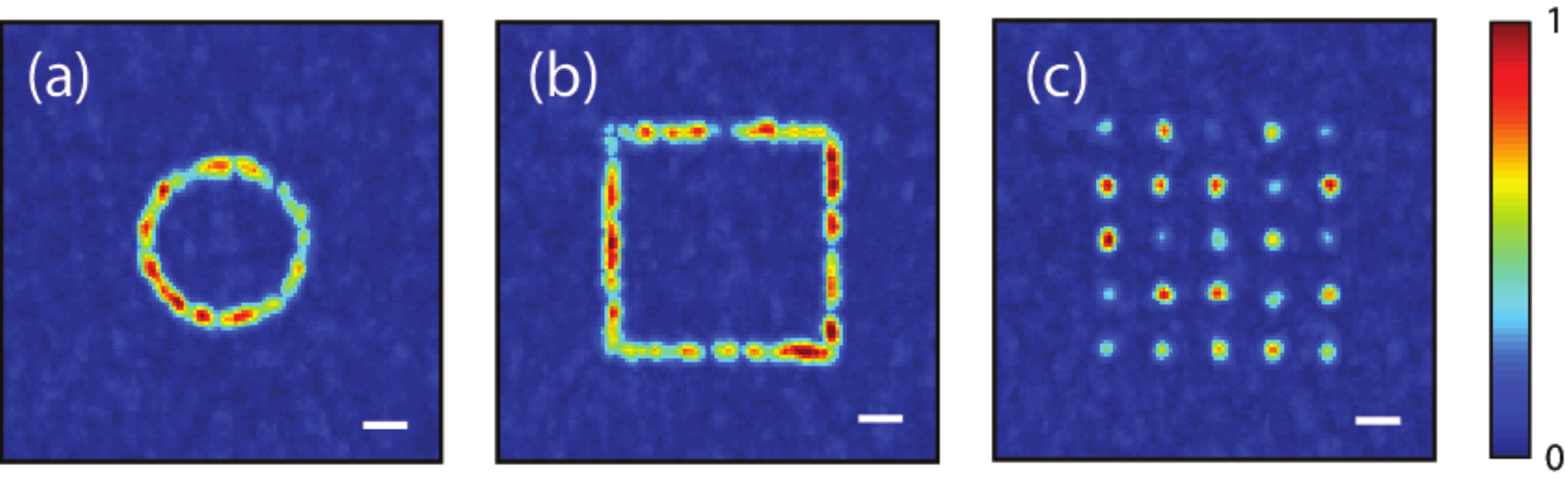}
\caption{Single-focus scanning allows time-averaged pattern projection through paper. Maximum intensity projections of patterns obtained when a single focus is scanned following (a) a circle (128 scanning points), (b) a square (256 scanning points), and (c) a grid of $5 \times 5$ points with the SLNN and paper as scattering material. Color bars: intensity (a.u.) normalized for each image. Scale bars:$2.7\,\rm{\mu m}$}
\label{sfig_paper}
\end{figure*}

\subsection*{Light control over different fields of view}
\begin{figure*}[hb]
\centering
\includegraphics[width=0.85\linewidth]{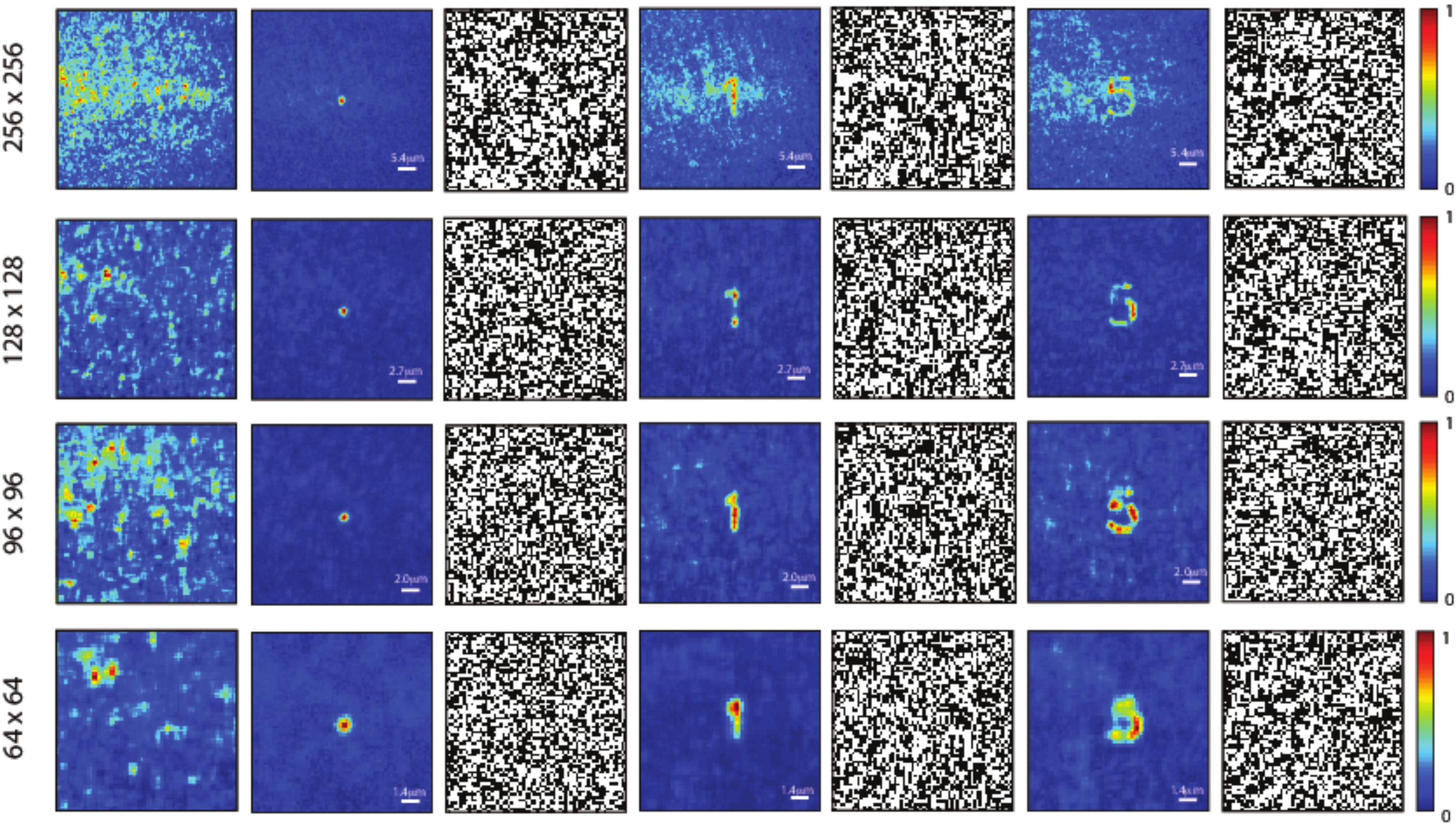}
\caption{Light control over different fields of view. Normalized transverse intensity distributions of an example of speckle pattern (first column), a single focus (second column), the number ``1" (fourth column), and the number ``5" (6th column) for different fields of view: $256 \times 256$ pixels (first row), $128 \times 128$ pixels (second row), $96 \times 96$ pixels (third row), and $64 \times 64$ pixels (last row). Columns number 3, 5, and 7 are the actual DMD patterns used to generate the light distributions from columns number 2, 4, and 6, respectively. 
Color bars: intensity (a.u.) normalized for each image.}
\label{sfig2}
\end{figure*}

In the main text we showed the ability of NNs to shape light through disordered media within a field of view of $20\times 20\,\rm{\mu m^2}$ imaged onto $96\times 96$ pixels on the CCD camera. However, the presented method performs similarly for other fields of view, as shown in Fig.~\ref{sfig2} for the case of the glass diffuser, with $64 \times 64$ macropixels displayed on the DMD controlled with the SLNN. In all cases, the experimental setup is the same as the one shown in the main text and the only difference is the resolution of the images captured with the CCD camera and used to train the SLNN. Note that the illumination resolution additionally impacts the quality of light shaping through the diffuser (in agreement with previous reports, see e.g. Ref.~\cite{2007:ol:vellekoop}).

\begin{figure*}[h!]
\centering
\includegraphics[width=0.85\linewidth]{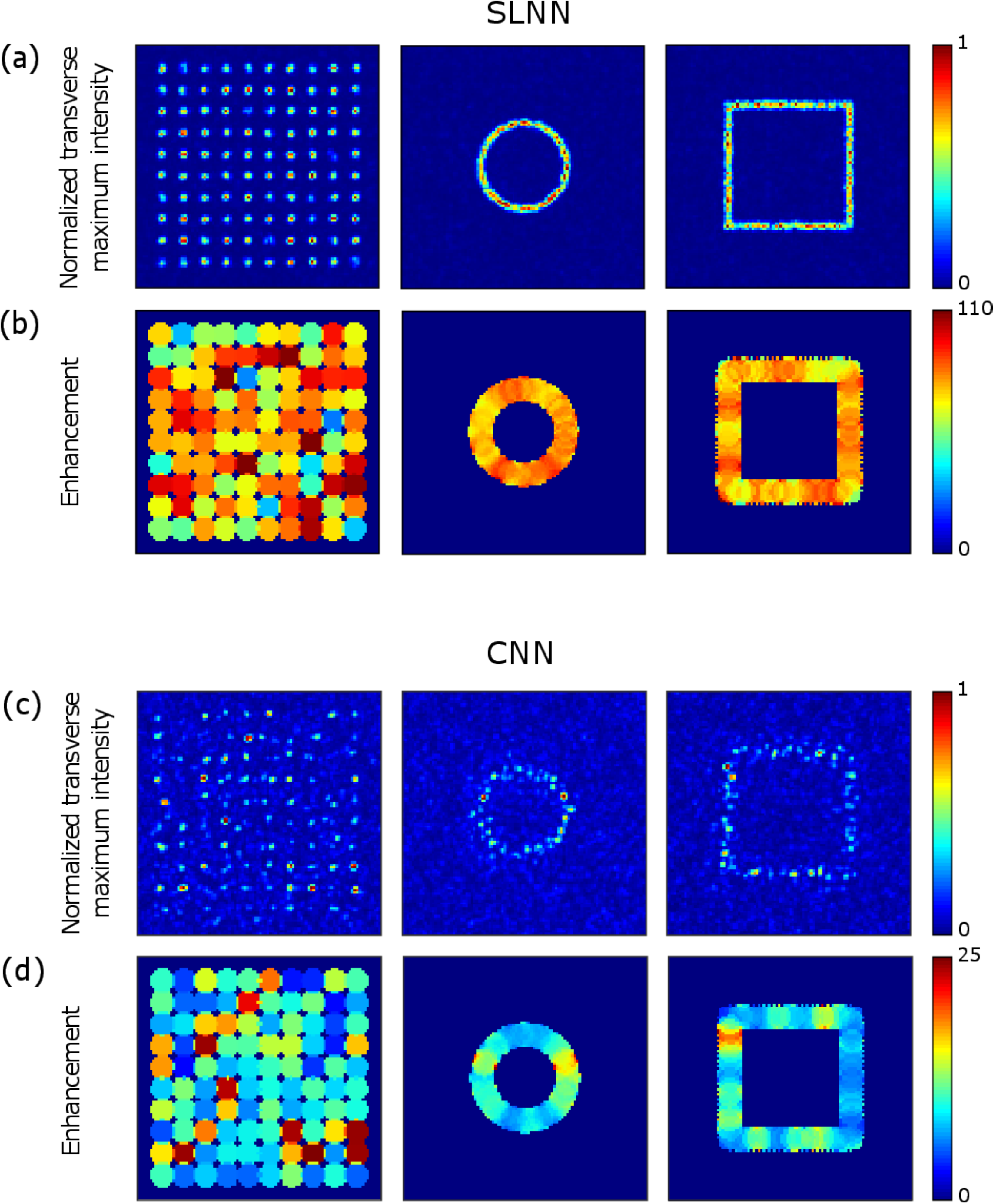}
\caption{Scanning and corresponding enhancement for focusing across paper with the SLNN (a, b) and the CNN (c, d) with the 40X objective. (a) shows different scan patterns across the entire field of view (grid, circle and square) for the SLNN, and (b) shows the corresponding enhancement for each point in (a). (c) Same as (a) for the CNN. (d) same as (b) for the CNN. The width and height of the full field of view is $51\,\rm{\mu m}$, the single pixel width and hight is $0.53\,\rm{\mu m}$. Note the different scale bars for the enhancement for the SLNN and CNN.}
\label{sfig_cnn_slnn}
\end{figure*}

\subsection*{Pattern generation by scanning a focus at high frequency}

Since neural networks can generate single foci with high fidelity, one can take advantage of the fast operation of the DMD ($22.7\,\rm{kHz}$) to obtain any transverse intensity distribution after the scatter by scanning a single spot (or multiple spots) at high speed. This is shown in \textcolor{blue}{Visualization} 1, \textcolor{blue}{Visualization 2}, \textcolor{blue}{Visualization 3}, \textcolor{blue}{Visualization 4}, \textcolor{blue}{Visualization 5}, \textcolor{blue}{Visualization 6}, \textcolor{blue}{Visualization 7}, \textcolor{blue}{Visualization 8}, \textcolor{blue}{Visualization 9}, \textcolor{blue}{Visualization 10}, \textcolor{blue}{Visualization 11}, \textcolor{blue}{Visualization 12} for a single focus tracing out a circle (\textcolor{blue}{Visualization 1}, \textcolor{blue}{Visualization 4}, \textcolor{blue}{Visualization 7}, \textcolor{blue}{Visualization 10}), a square (\textcolor{blue}{Visualization 2}, \textcolor{blue}{Visualization 5}, \textcolor{blue}{Visualization 8}, \textcolor{blue}{Visualization 3}, \textcolor{blue}{Visualization 6}, \textcolor{blue}{Visualization 9}, \textcolor{blue}{Visualization 12}) consisting of 128/96, 256/256, and 25/25 scanning positions, respectively, through a glass diffuser both for the SLNN (\textcolor{blue}{Visualization 1}, \textcolor{blue}{Visualization 2}, \textcolor{blue}{Visualization 3}) and the CNN (\textcolor{blue}{Visualization 4}, \textcolor{blue}{Visualization 5}, \textcolor{blue}{Visualization 6}); a multimode optical fiber (\textcolor{blue}{Visualization 7}, \textcolor{blue}{Visualization 8}, \textcolor{blue}{Visualization 9}), and a piece of paper (\textcolor{blue}{Visualization 10}, \textcolor{blue}{Visualization 11}, \textcolor{blue}{Visualization 12}). These sequences were recorded at a speed of $500\,\rm{Hz}$ (the fastest allowed by our CCD camera). Note that for the DMD operating at full speed, i.e. without restrictions imposed by the camera, a sequence composed of 96 focus positions projecting a certain pattern will result in a pattern projection frame rate higher than $200\,\rm{Hz}$. 

To quantify the focusing accuracy of the different network architectures presented in Fig. 11 we calculated the average distance between the target focus position and the measured focus position (which was measured by finding the focus centroid by fitting it with a Gaussian distributions). The values obtained for the  SLNN were 0.63 pixels (standard deviation 0.20) and for the CNN 0.99 pixels, (standard deviation 0.60), with a pixel size of $0.53\,\rm{\mu m}$.

\subsection*{Neural network design and performance}
\label{nns}
We use the Keras library \cite{chollet2015keras} with the TensorFlow \cite{tensorflow2015-whitepaper} back-end for GPU-accelerated neural network training. The networks are trained to map grayscale speckle images to the corresponding binary illumination patterns with a subset of the total dataset of image pairs (8000 pairs in our case) and tested on the previously not introduced data (the remaining 2000 pairs). Once the network is trained, we input the desired PSF and the output binary map is uploaded to the DMD for light control through the diffuser or fiber. 

\begin{figure*}[h!]
\centering
\includegraphics[width=0.60\linewidth]{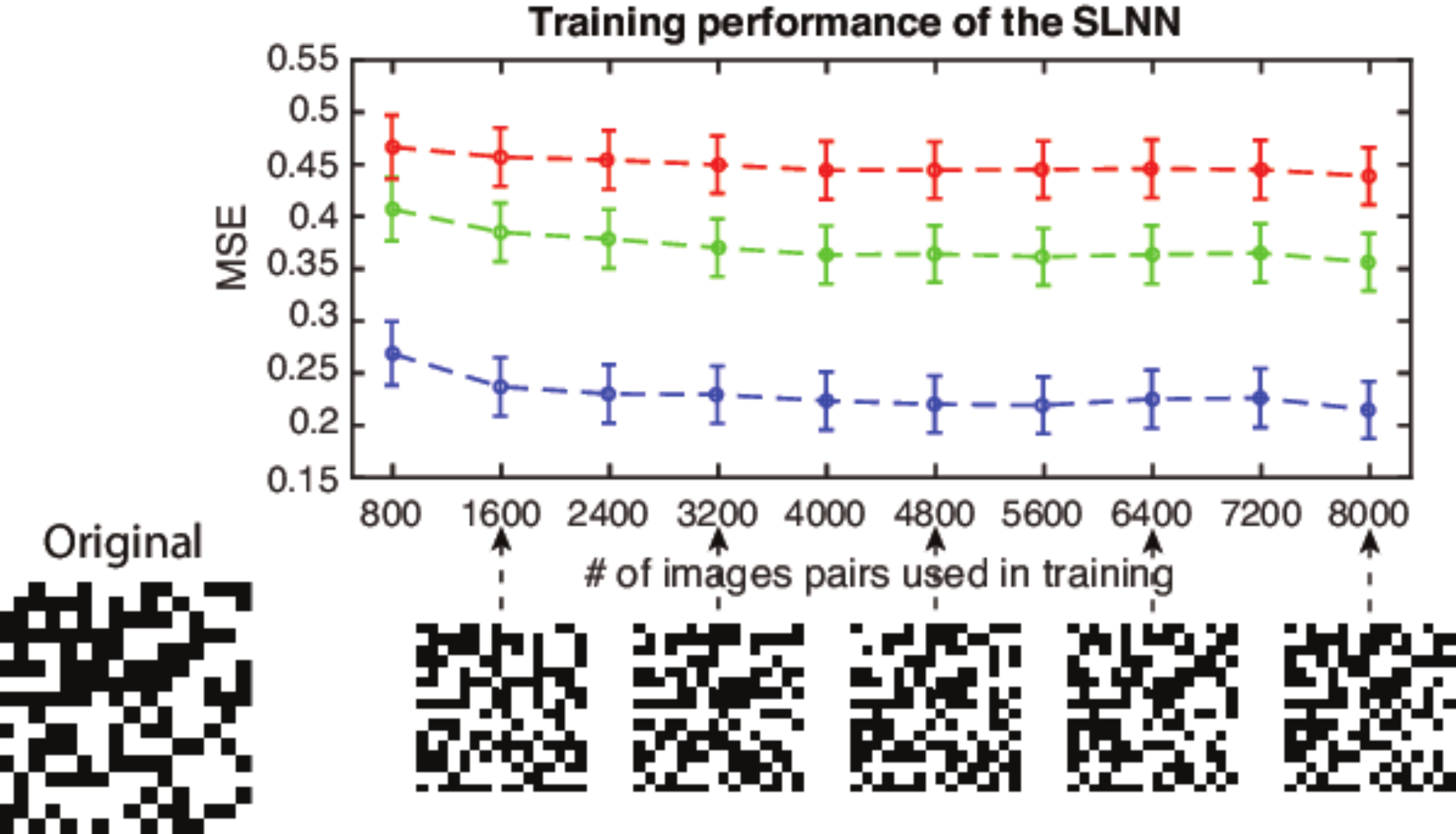}
\caption{Training performance of SLNN. Mean-square error (MSE) between the predicted and original illumination patterns for the single-layer neural network for different sizes of the dataset. Red-, green-, and blue-dashed curves correspond to illumination sizes of $64 \times 64$, $32 \times 32$, and $16 \times 16$, respectively. Insets below show the the predicted illumination at different stages of the training for the $16 \times 16$ case.}
\label{sfig3}
\end{figure*}

\begin{figure*}[h!]
\centering
\includegraphics[width=0.80\linewidth]{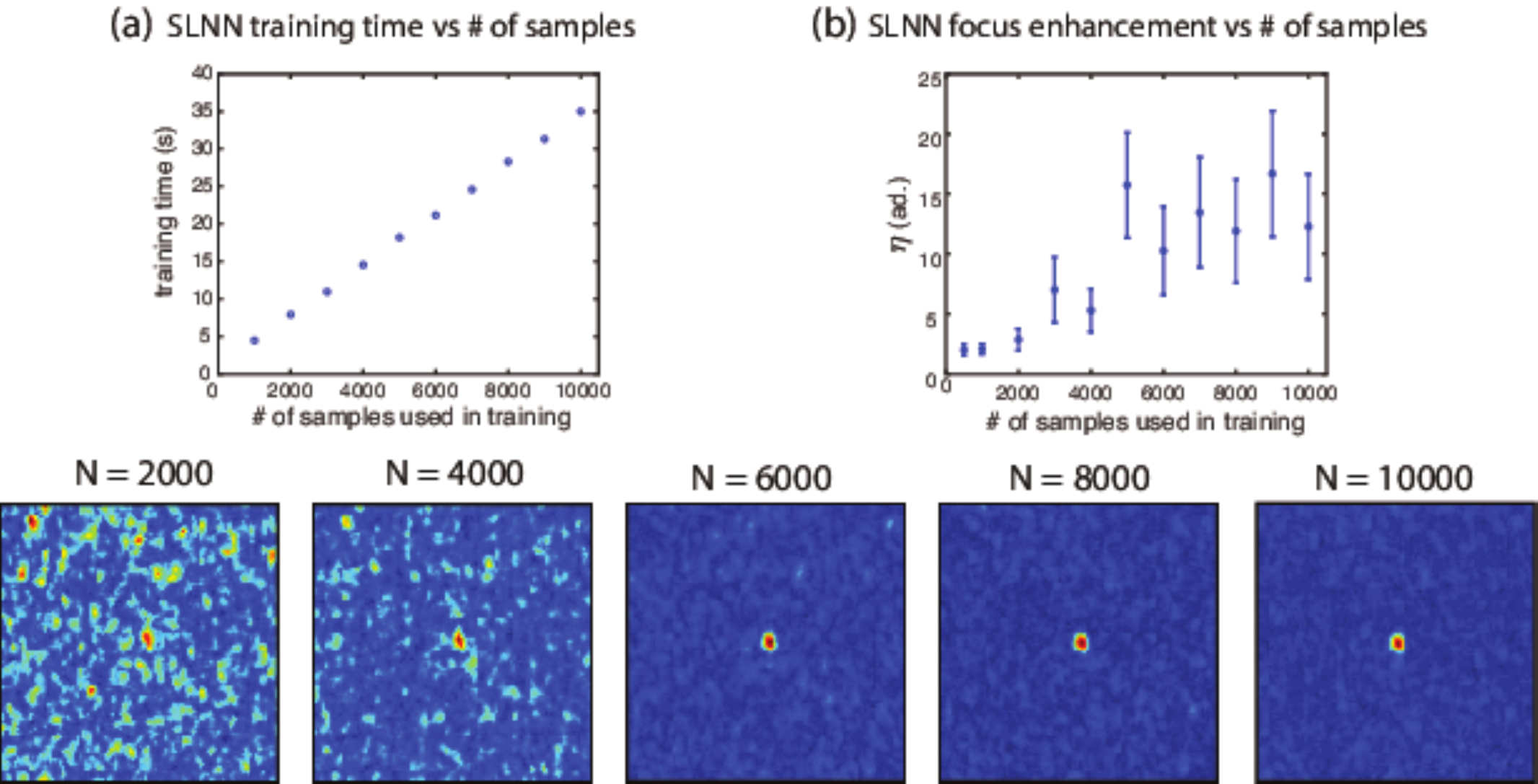}
\caption{Speed of SLNN training and quality of focus. (a) Training time of the SLNN and (b) enhancement $\eta$ of the generated foci with paper as scattering sample for different values of the number of image pairs used during training. The enhancement is defined as $\eta \equiv I_{\rm{focus}}/\langle I_{\rm{speckle}} \rangle$, where $I_{\rm{focus}}$ is the intensity at the generated foci and $\langle I_{\rm{speckle}} \rangle$ is the mean value of the background speckle \cite{2011:oe:mosk}. Image in the lower part of the figure are examples of focusing with $N$ samples used for training (with the number $N$ of samples used for training at the top of each image.}
\label{sfig4}
\end{figure*}
The SLNN we used is a single-layer perceptron, which is a network consisting of one fully connected layer followed by a non-linear activation function bounding the output to the 0-1 range. In principle, it can be represented as a matrix dot product, with bias addition and a sigmoid function applied element-wise to the resulting vector. We found that with the activation function applied per each individual element the model is prone to over-fitting and does not make good generalizations. As a solution, we replaced the nonlinear activation function with a binarization function with a threshold common for the whole predicted pattern (mean value of the prediction) which results in a more robust model with better focus enhancement and faster training. The training time depends on the number of images used (8000 in our case), the batch size (number of images taken for each iteration of the training algorithm, 150 in our case), and the number of epochs (up to 20 for the results presented here, see Figs.~\ref{sfig3} and \ref{sfig3} for SLNN training performance). With these parameters the single-layer perceptron requires less than 35 seconds for training, while the predicted patterns take about $1\,\rm{s}$ to be calculated. However, we have verified that lower training times with acceptable enhancement can be obtained by reducing the number of image pairs used in the training. For an analysis on how the training times and enhancement of the focus depend with the number of images pairs used in the training, see Fig.~\ref{sfig4}. 

For exploring functional relations between transmitted and reflected speckle patterns we concatenated two SLNNs similar to the one described above. The first SLNN (SLNN1) uses the transmitted speckle as input and connects it to the reflected speckle pattern (output) through a single fully-connected layer (without binarization). The second SLNN (SLNN2) connects the reflected speckle (input) to the illumination patterns (output) also through a fully connected layer including the mean threshold binarization. 
Both sets of speckle patterns can be generated with independent checkerboard illumination patterns. For training of these two SLNNs we used the same number of images, batch size, and number of epochs as for the SLNN discussed above, with similar performance. The desired illumination in transmission is finally fed into SLNN1, which predicts a speckle pattern as output, which in turn serves as input for SLNN2. The output of SLNN2 is a binary illumination pattern that is sent to the DMD in order to experimentally obtain the desired illumination.

In perceptron-like models a single fully-connected layer contains a large number of parameters (the product of input and output vector dimensions) which makes these models more demanding to train as the resolution of the illumination and speckle images and the memory demand increase. CNNs can efficiently reduce the number of trainable parameters and we first used a model with three convolutional layers with 48 ($9\times9$), 24 ($5 \times 5$) and 12 ($3\times3$) filters respectively, each succeeded by rectified linear unit (ReLU) activation and ($2 \times 2$) max pooling operation, followed by a fully connected layer with 0.25 dropout rate (see below and Fig. 11 for a different CNN architecture). This configuration achieves a performance similar to the SLNN in controlling a single focal spot while having 20\%  of the SLNN's number of parameters. 

As any deeper network, the CNN requires longer training time and a more extensive dataset. A workaround is offered by the fine tuning technique: the convolutional layers are pretrained separately on a dataset of 40000 speckle images in an autoencoder. An autoencoder is a network trained to map its input to itself, however it contains a bottleneck - a lower dimensional middle layer (latent space) where a compressed representation of the data is learned. Our autoencoder has three convolutional layers as needed for the proposed CNN model and a symmetrical deconvolutional decoder. 
The training time largely varies with dataset size and speckle image resolution, and it is best to provide as much data as possible. Good training results were achieved after 20 minutes of training with 40000 samples sized $256 \times 256$. 

For the results presented in Fig. \ref{sfig_cnn_slnn}, a different CNN structure was used: two convolutional layers of 48 ($7\times7$), and 24 ($5 \times 5$) filters, both succeeded by rectified linear unit (ReLU) activation, ($2 \times 2$) max pooling and batch normalization and a 0.3 dropout used in the training. The subsequent fully-connected layer used sigmoid activation. This network could be trained from scratch in 150 seconds (25 epochs of 10000 samples dataset, with batch size of 100). 
Additionally, the input scaling used in the illumination prediction was adapted. The CNN is trained on images produced by random illumination patterns, which contain speckles equally distributed across the whole field of view. The image of the focus which serves as the input in the prediction stage is however significantly different: it has zeros in most parts except for a Gaussian peak. This input results in lower activation values in the intermediate layers compared to random speckle input images, and as a result the output illumination is not binary, but rather continuous-valued, because the sigmoid activation units of the output layer do not receive input of sufficient magnitude to saturate to either 0 or 1 values. To overcome this, the focus input is scaled by empirically adjusting the value of $10^5$ to force the CNN to output binary patterns.

\subsection*{Training performance of SLNN}
Although neural networks are capable of predicting the illumination patterns necessary for light control through scattering media, there are multiple variables affecting the efficiency of the training. Here, we discuss the impact of the size of the dataset used for training. In Fig. \ref{sfig3} we plot the mean-square error (MSE) between the predicted illuminations after training the single-layer neural network and a set of 100 original illuminations that were not included in the training. The analysis is performed for different sizes of the dataset (ranging from 800 pairs to 8000 pairs) and different illumination sizes: $64 \times 64$ (red), $32 \times 32$ (green), and $16 \times 16$ (blue). As an illustration of the training performance, we have included an example of original illumination and corresponding predictions obtained when 1600, 3200, 4800, 6200, and 8000 pairs are used ($16 \times 16$ case). As expected, the lower the size of the illumination pattern, the higher the MSE. 

In Fig. \ref{sfig4} we further explore the training performance of the SLNN as the size of the dataset used in the training varies, in this case in  terms of Fig. \ref{sfig4}(a) training time, and Fig. \ref{sfig4}(b) enhancement of the generated foci. While the training time shows a linearly increasing tendency, the enhancement tends to saturate beyond a certain number of samples (typically for $N > 5000$).

\section*{Acknowledgements}
We thank Andres Flores for Python support and Bernd Scheiding for electronics support. We gratefully acknowledge the support of NVIDIA Corporation with the donation of the Titan Xp GPU used for this research. 
This work was supported by the Max Planck Society and the Center of Advanced European Studies and Research (caesar).

\section*{Author Contributions}
A.T. and I.V. performed the experiments and analyzed the data. I.V. designed the neural network analysis. A.T., I.V., and J.D.S. designed the project and wrote the manuscript. 


\section*{Competing Interests}
The authors declare no competing financial interests.


\begin{thebibliography}{10}
\expandafter\ifx\csname url\endcsname\relax
  \def\url#1{\texttt{#1}}\fi
\expandafter\ifx\csname urlprefix\endcsname\relax\def\urlprefix{URL }\fi
\providecommand{\bibinfo}[2]{#2}
\providecommand{\eprint}[2][]{\url{#2}}

\bibitem{2017:rmp:rotter}
\bibinfo{author}{Rotter, S.} \& \bibinfo{author}{Gigan, S.}
\newblock \bibinfo{title}{Light fields in complex media: Mesoscopic scattering
  meets wave control}.
\newblock \emph{\bibinfo{journal}{Reviews of Modern Physics}}
  \textbf{\bibinfo{volume}{89}}, \bibinfo{pages}{015005}
  (\bibinfo{year}{2017}).

\bibitem{2017:natmeth:naji}
\bibinfo{author}{Ji, N.}
\newblock \bibinfo{title}{Adaptive optical fluorescence microscopy}.
\newblock \emph{\bibinfo{journal}{Nature Methods}}
  \textbf{\bibinfo{volume}{14}}, \bibinfo{pages}{374--380}
  (\bibinfo{year}{2017}).

\bibitem{2012:natcommun:dholakia}
\bibinfo{author}{Čižmár, T.} \& \bibinfo{author}{Dholakia, K.}
\newblock \bibinfo{title}{Exploiting multimode waveguides for pure fibre-based
  imaging}.
\newblock \emph{\bibinfo{journal}{Nature Communications}}
  \textbf{\bibinfo{volume}{3}}, \bibinfo{pages}{1027} (\bibinfo{year}{2012}).

\bibitem{2010:natphoton:dholakia}
\bibinfo{author}{Čižmár, T.}, \bibinfo{author}{Mazilu, M.} \&
  \bibinfo{author}{Dholakia, K.}
\newblock \bibinfo{title}{In situ wavefront correction and its application to
  micromanipulation}.
\newblock \emph{\bibinfo{journal}{Nature Photonics}}
  \textbf{\bibinfo{volume}{3}}, \bibinfo{pages}{388--394}
  (\bibinfo{year}{2010}).

\bibitem{2017:sa:yang}
\bibinfo{author}{\color{\changeshighlight} H.~W.~Ruan} \emph{et~al.}
\newblock \bibinfo{title}{Deep tissue optical focusing and optogenetic
  modulation with time-reversed ultrasonically encoded light}.
\newblock \emph{\bibinfo{journal}{Science Advances}}
  \textbf{\bibinfo{volume}{3}}, \bibinfo{pages}{eaao5520} (\bibinfo{year}{2017
  \color{black}}).

\bibitem{2012:prl:choi}
\bibinfo{author}{Choi, Y.} \emph{et~al.}
\newblock \bibinfo{title}{Scanner-free and wide-field endoscopic imaging by
  using a single multimode optical fiber}.
\newblock \emph{\bibinfo{journal}{Physical Review Letters}}
  \textbf{\bibinfo{volume}{109}}, \bibinfo{pages}{203901}
  (\bibinfo{year}{2012}).

\bibitem{2016:aop:forbes}
\bibinfo{author}{Forbes, A.}, \bibinfo{author}{Dudley, A.} \&
  \bibinfo{author}{McLaren, M.}
\newblock \bibinfo{title}{Creation and detection of optical modes with spatial
  light modulators}.
\newblock \emph{\bibinfo{journal}{Adv. Opt. Photon}}
  \textbf{\bibinfo{volume}{8}}, \bibinfo{pages}{200--227}
  (\bibinfo{year}{2016}).

\bibitem{2017:oe:cizmar}
\bibinfo{author}{Turtaev, S.} \emph{et~al.}
\newblock \bibinfo{title}{Comparison of nematic liquid-crystal and dmd based
  spatial light modulation in complex photonics}.
\newblock \emph{\bibinfo{journal}{Optics Express}}
  \textbf{\bibinfo{volume}{25}}, \bibinfo{pages}{29874--29884}
  (\bibinfo{year}{2017}).

\bibitem{2007:ol:vellekoop}
\bibinfo{author}{Vellekoop, I.~M.} \& \bibinfo{author}{Mosk, A.~P.}
\newblock \bibinfo{title}{Focusing coherent light through opaque strongly
  scattering media}.
\newblock \emph{\bibinfo{journal}{Optics Letters}}
  \textbf{\bibinfo{volume}{32}}, \bibinfo{pages}{2309--2311}
  (\bibinfo{year}{2007}).

\bibitem{2010:prl:gigan}
\bibinfo{author}{Popoff, S.~M.} \emph{et~al.}
\newblock \bibinfo{title}{Measuring the transmission matrix in optics: An
  approach to the study and control of light propagation in disordered media}.
\newblock \emph{\bibinfo{journal}{Physical Review Letters}}
  \textbf{\bibinfo{volume}{104}}, \bibinfo{pages}{100601}
  (\bibinfo{year}{2010}).

\bibitem{2012:oe:piestun}
\bibinfo{author}{Conkey, D.~B.}, \bibinfo{author}{Caravaca-Aguirre, A.~M.} \&
  \bibinfo{author}{Piestun, R.}
\newblock \bibinfo{title}{High-speed scattering medium characterization with
  application to focusing light through turbid media}.
\newblock \emph{\bibinfo{journal}{Optics Express}}
  \textbf{\bibinfo{volume}{20}}, \bibinfo{pages}{1733--1740}
  (\bibinfo{year}{2012}).

\bibitem{2011:oe:dileonardo}
\bibinfo{author}{Leonardo, R.~D.} \& \bibinfo{author}{Bianchi, S.}
\newblock \bibinfo{title}{Hologram transmission through multi-mode optical
  fibers}.
\newblock \emph{\bibinfo{journal}{Optics Express}}
  \textbf{\bibinfo{volume}{19}}, \bibinfo{pages}{247--254}
  (\bibinfo{year}{2011}).

\bibitem{2014:ol:choi}
\bibinfo{author}{Kim, D.} \emph{et~al.}
\newblock \bibinfo{title}{Toward a miniature endomicroscope: pixelation-free
  and diffraction-limited imaging through a fiber bundle}.
\newblock \emph{\bibinfo{journal}{Optics Express}}
  \textbf{\bibinfo{volume}{39}}, \bibinfo{pages}{1291--1294}
  (\bibinfo{year}{2014}).

\bibitem{2017:sciadv:yang}
\bibinfo{author}{Ruan, H.} \emph{et~al.}
\newblock \bibinfo{title}{Deep tissue optical focusing and optogenetic
  modulation with time-reversed ultrasonically encoded light}.
\newblock \emph{\bibinfo{journal}{Science Advances}}
  \textbf{\bibinfo{volume}{3}}, \bibinfo{pages}{eaao5520}
  (\bibinfo{year}{2017}).

\bibitem{2015:scirep:park}
\bibinfo{author}{Yoon, J.} \emph{et~al.}
\newblock \bibinfo{title}{Optogenetic control of cell signaling pathway through
  scattering skull using wavefront shaping}.
\newblock \emph{\bibinfo{journal}{Scientific Reports}}
  \textbf{\bibinfo{volume}{3}}, \bibinfo{pages}{13289} (\bibinfo{year}{2015}).

\bibitem{2015:optica:yang}
\bibinfo{author}{Wang, D.} \emph{et~al.}
\newblock \bibinfo{title}{Focusing through dynamic tissue with millisecond
  digital optical phase conjugation}.
\newblock \emph{\bibinfo{journal}{Optica}} \textbf{\bibinfo{volume}{2}},
  \bibinfo{pages}{728--735} (\bibinfo{year}{2015}).

\bibitem{2015:natcommun:wang}
\bibinfo{author}{Liu, Y.}, \bibinfo{author}{Lai, P.}, \bibinfo{author}{Ma, C.},
  \bibinfo{author}{Xu, X.} \& \bibinfo{author}{andLihong V.~Wang, A. A.~G.}
\newblock \bibinfo{title}{Optical focusing deep inside dynamic scattering media
  with near-infrared time-reversed ultrasonically encoded (true) light}.
\newblock \emph{\bibinfo{journal}{Nature Communications}}
  \textbf{\bibinfo{volume}{6}}, \bibinfo{pages}{5904} (\bibinfo{year}{2015}).

\bibitem{2015:oe:vellekoop}
\bibinfo{author}{Vellekoop, I.~M.}
\newblock \bibinfo{title}{Feedback-based wavefront shaping}.
\newblock \emph{\bibinfo{journal}{Optics Express}}
  \textbf{\bibinfo{volume}{23}}, \bibinfo{pages}{12189--12206}
  (\bibinfo{year}{2015}).

\bibitem{2015:natphton:horstmeyer}
\bibinfo{author}{Horstmeyer, R.}, \bibinfo{author}{Ruan, H.} \&
  \bibinfo{author}{Yang, C.}
\newblock \bibinfo{title}{Guidestar-assisted wavefront-shaping methods for
  focusing light into biological tissue}.
\newblock \emph{\bibinfo{journal}{Nature Photonics}}
  \textbf{\bibinfo{volume}{9}}, \bibinfo{pages}{563--571}
  (\bibinfo{year}{2015}).

\bibitem{2012:natphoton:mosk}
\bibinfo{author}{Mosk, A.~P.}, \bibinfo{author}{Lagendijk, A.},
  \bibinfo{author}{Lerosey, G.} \& \bibinfo{author}{Fink, M.}
\newblock \bibinfo{title}{Controlling waves in space and time for imaging and
  focusing in complex media}.
\newblock \emph{\bibinfo{journal}{Nature Photonics}}
  \textbf{\bibinfo{volume}{6}}, \bibinfo{pages}{283--292}
  (\bibinfo{year}{2012}).

\bibitem{2012:PNAS:cui}
\bibinfo{author}{Tanga, J.}, \bibinfo{author}{Germain, R.~N.},  \&
  \bibinfo{author}{Cui, M.}
\newblock \bibinfo{title}{Superpenetration optical microscopy by iterative
  multiphoton adaptive compensation technique}.
\newblock \emph{\bibinfo{journal}{Proceedings of the National Academy of
  Sciences}} \textbf{\bibinfo{volume}{129}}, \bibinfo{pages}{8434--8439}
  (\bibinfo{year}{2012}).

\bibitem{2017:ol:gigan}
\bibinfo{author}{Blochet, B.}, \bibinfo{author}{Bourdieu, L.} \&
  \bibinfo{author}{Gigan, S.}
\newblock \bibinfo{title}{Focusing light through dynamical samples using fast
  continuous wavefront optimization}.
\newblock \emph{\bibinfo{journal}{Optics Letters}}
  \textbf{\bibinfo{volume}{42}}, \bibinfo{pages}{4994--4997}
  (\bibinfo{year}{2017}).

\bibitem{2010:oe:cui}
\bibinfo{author}{Cui, M.} \& \bibinfo{author}{Yang, C.}
\newblock \bibinfo{title}{Implementation of a digital optical phase conjugation
  system and its application to study the robustness of turbidity suppression
  by phase conjugation}.
\newblock \emph{\bibinfo{journal}{Optics Express}}
  \textbf{\bibinfo{volume}{18}}, \bibinfo{pages}{3444--3455}
  (\bibinfo{year}{2010}).

\bibitem{2010:oe:psaltis}
\bibinfo{author}{Hsieh, C.-L.}, \bibinfo{author}{Pu, Y.},
  \bibinfo{author}{Grange, R.}, \bibinfo{author}{Laporte, G.} \&
  \bibinfo{author}{Psaltis, D.}
\newblock \bibinfo{title}{Imaging through turbid layers by scanning the phase
  conjugated second harmonic radiation from a nanoparticle}.
\newblock \emph{\bibinfo{journal}{Optics Express}}
  \textbf{\bibinfo{volume}{18}}, \bibinfo{pages}{20723--20731}
  (\bibinfo{year}{2010}).

\bibitem{2012:natcommun:judkewitz}
\bibinfo{author}{Wang, Y.~M.}, \bibinfo{author}{Judkewitz, B.},
  \bibinfo{author}{DiMarzio, C.~A.} \& \bibinfo{author}{Yang, C.}
\newblock \bibinfo{title}{Deep-tissue focal fluorescence imaging with digitally
  time-reversed ultrasound-encoded light}.
\newblock \emph{\bibinfo{journal}{Nature Communications}}
  \textbf{\bibinfo{volume}{3}}, \bibinfo{pages}{928} (\bibinfo{year}{2012}).

\bibitem{2013:scirep:yaqoob}
\bibinfo{author}{Hillman, T.~R.} \emph{et~al.}
\newblock \bibinfo{title}{Digital optical phase conjugation for delivering
  two-dimensional images through turbid media}.
\newblock \emph{\bibinfo{journal}{Scientific Reports}}
  \textbf{\bibinfo{volume}{3}}, \bibinfo{pages}{1909} (\bibinfo{year}{2013}).

\bibitem{2017:optica:wang}
\bibinfo{author}{Liu, Y.}, \bibinfo{author}{Ma, C.}, \bibinfo{author}{Shen,
  Y.}, \bibinfo{author}{Shi, J.} \& \bibinfo{author}{Wang, L.~V.}
\newblock \bibinfo{title}{Focusing light inside dynamic scattering media with
  millisecond digital optical phase conjugation}.
\newblock \emph{\bibinfo{journal}{Optica}} \textbf{\bibinfo{volume}{4}},
  \bibinfo{pages}{280--288} (\bibinfo{year}{2017}).

\bibitem{2010:natcommun:gigan}
\bibinfo{author}{Popoff, S.}, \bibinfo{author}{Lerosey, G.},
  \bibinfo{author}{Fink, M.}, \bibinfo{author}{Boccara, A.~C.} \&
  \bibinfo{author}{Gigan, S.}
\newblock \bibinfo{title}{Image transmission through an opaque material}.
\newblock \emph{\bibinfo{journal}{Nature Communications}}
  \textbf{\bibinfo{volume}{1}}, \bibinfo{pages}{81} (\bibinfo{year}{2010}).

\bibitem{2017:optica:gigan}
\bibinfo{author}{Boniface, A.}, \bibinfo{author}{Mounaix, M.},
  \bibinfo{author}{Blochet, B.}, \bibinfo{author}{Piestun, R.} \&
  \bibinfo{author}{Gigan, S.}
\newblock \bibinfo{title}{Transmission-matrix-based point-spread-function
  engineering through a complex medium}.
\newblock \emph{\bibinfo{journal}{Optica}} \textbf{\bibinfo{volume}{4}},
  \bibinfo{pages}{54--59} (\bibinfo{year}{2017}).

\bibitem{2015:oe:Dremeau}
\bibinfo{author}{Drémeau, A.} \emph{et~al.}
\newblock \bibinfo{title}{Reference-less measurement of the transmission matrix
  of a highly scattering material using a dmd and phase retrieval techniques}.
\newblock \emph{\bibinfo{journal}{Optics Express}}
  \textbf{\bibinfo{volume}{23}}, \bibinfo{pages}{11898--11911}
  (\bibinfo{year}{2015}).

\bibitem{2017:ieee:metzler}
\bibinfo{author}{Metzler, C.~A.} \emph{et~al.}
\newblock \bibinfo{title}{Coherent inverse scattering via transmission
  matrices: Efficient phase retrieval algorithms and a public dataset}.
\newblock In \emph{\bibinfo{booktitle}{2017 IEEE International Conference on
  Computational Photography (ICCP)}}, \bibinfo{pages}{1--16}
  (\bibinfo{year}{2017}).

\bibitem{2015:science:mitchell}
\bibinfo{author}{Jordan, M.~I.} \& \bibinfo{author}{Mitchell, T.~M.}
\newblock \bibinfo{title}{Machine learning: Trends, perspectives, and
  prospects}.
\newblock \emph{\bibinfo{journal}{Science}} \textbf{\bibinfo{volume}{349}},
  \bibinfo{pages}{255--260} (\bibinfo{year}{2015}).

\bibitem{2015:nature:hinton}
\bibinfo{author}{LeCun, Y.}, \bibinfo{author}{Bengio, Y.} \&
  \bibinfo{author}{Hinton, G.}
\newblock \bibinfo{title}{Deep learning}.
\newblock \emph{\bibinfo{journal}{Nature}} \textbf{\bibinfo{volume}{521}},
  \bibinfo{pages}{436--444} (\bibinfo{year}{2015}).

\bibitem{2015:nature:waller}
\bibinfo{author}{Waller, L.} \& \bibinfo{author}{Tian, L.}
\newblock \bibinfo{title}{Machine learning for 3d microscopy}.
\newblock \emph{\bibinfo{journal}{Nature}} \textbf{\bibinfo{volume}{523}},
  \bibinfo{pages}{416--417} (\bibinfo{year}{2015}).

\bibitem{2018:prl:Deans}
\bibinfo{author}{Deans, C.}, \bibinfo{author}{Griffin, L.~D.},
  \bibinfo{author}{Marmugi, L.},  \& \bibinfo{author}{Renzoni, F.}
\newblock \bibinfo{title}{Machine learning based localization and
  classification with atomic magnetometers}.
\newblock \emph{\bibinfo{journal}{Physical Review Letters}}
  \textbf{\bibinfo{volume}{120}}, \bibinfo{pages}{033204}
  (\bibinfo{year}{2018}).

\bibitem{2015:optica:Kamilov}
\bibinfo{author}{Kamilov, U.~S.} \emph{et~al.}
\newblock \bibinfo{title}{Learning approach to optical tomography}.
\newblock \emph{\bibinfo{journal}{Optica}} \textbf{\bibinfo{volume}{2}},
  \bibinfo{pages}{517--522} (\bibinfo{year}{2015}).

\bibitem{2017:optica_lensless:barbastathis}
\bibinfo{author}{Sinha, A.}, \bibinfo{author}{Li, J. L.~S.} \&
  \bibinfo{author}{Barbastathis, G.}
\newblock \bibinfo{title}{Lensless computational imaging through deep
  learning}.
\newblock \emph{\bibinfo{journal}{Optica}} \textbf{\bibinfo{volume}{4}},
  \bibinfo{pages}{1117--1125} (\bibinfo{year}{2017}).

\bibitem{2017:optica:ozcan}
\bibinfo{author}{Rivenson, Y.} \emph{et~al.}
\newblock \bibinfo{title}{Deep learning microscopy}.
\newblock \emph{\bibinfo{journal}{Optica}} \textbf{\bibinfo{volume}{4}},
  \bibinfo{pages}{1437--1443} (\bibinfo{year}{2017}).

\bibitem{2018:arxiv:shechtman}
\bibinfo{author}{Nehme, E.}, \bibinfo{author}{Weiss, L.~E.},
  \bibinfo{author}{Michaeli, T.},  \& \bibinfo{author}{Shechtman, Y.}
\newblock \bibinfo{title}{Deep-storm: Super resolution single molecule
  microscopy by deep learning}.
\newblock \emph{\bibinfo{journal}{arXiv preprint}}
  \textbf{\bibinfo{volume}{arXiv:1801.09631v1}} (\bibinfo{year}{2018}).

\bibitem{2018:optica:psaltis}
\bibinfo{author}{Borhani, N.}, \bibinfo{author}{Kakkava, E.},
  \bibinfo{author}{Moser, C.} \& \bibinfo{author}{Psaltis, D.}
\newblock \bibinfo{title}{Learning to see through multimode fibers}.
\newblock \emph{\bibinfo{journal}{Optica}} \textbf{\bibinfo{volume}{106}},
  \bibinfo{pages}{960--966} (\bibinfo{year}{2018}).

\bibitem{2012:oe_ga:piestun}
\bibinfo{author}{Conkey, D.~B.}, \bibinfo{author}{Brown, A.~N.},
  \bibinfo{author}{Caravaca-Aguirre, A.~M.} \& \bibinfo{author}{Piestun, R.}
\newblock \bibinfo{title}{Genetic algorithm optimization for focusing through
  turbid media in noisy environments}.
\newblock \emph{\bibinfo{journal}{Journal of Optics}}
  \textbf{\bibinfo{volume}{20}}, \bibinfo{pages}{4840--4849}
  (\bibinfo{year}{2012}).

\bibitem{2014:jo:kner}
\bibinfo{author}{Zhang, X.} \& \bibinfo{author}{Kner, P.}
\newblock \bibinfo{title}{Binary wavefront optimization using a genetic
  algorithm}.
\newblock \emph{\bibinfo{journal}{Journal of Optics}}
  \textbf{\bibinfo{volume}{17}}, \bibinfo{pages}{125704}
  (\bibinfo{year}{2014}).

\bibitem{2015:oe:kner}
\bibinfo{author}{Tehrani, K.~F.}, \bibinfo{author}{Xu, J.},
  \bibinfo{author}{Zhang, Y.}, \bibinfo{author}{Shen, P.} \&
  \bibinfo{author}{Kner, P.}
\newblock \bibinfo{title}{Adaptive optics stochastic optical reconstruction
  microscopy (ao-storm) using a genetic algorithm}.
\newblock \emph{\bibinfo{journal}{Optics Express}}
  \textbf{\bibinfo{volume}{23}}, \bibinfo{pages}{13677--13692}
  (\bibinfo{year}{2015}).

\bibitem{2017:jo:ding}
\bibinfo{author}{Zhang, B.} \emph{et~al.}
\newblock \bibinfo{title}{Focusing light through strongly scattering media
  using genetic algorithm with sbr discriminant}.
\newblock \emph{\bibinfo{journal}{Journal of Optics}}
  \textbf{\bibinfo{volume}{20}}, \bibinfo{pages}{025601}
  (\bibinfo{year}{2017}).

\bibitem{2015:oe:tanida}
\bibinfo{author}{Ando, T.}, \bibinfo{author}{Horisaki, R.} \&
  \bibinfo{author}{Tanida, J.}
\newblock \bibinfo{title}{Speckle-learning-based object recognition through
  scattering media}.
\newblock \emph{\bibinfo{journal}{Optics Express}}
  \textbf{\bibinfo{volume}{23}}, \bibinfo{pages}{33902--33910}
  (\bibinfo{year}{2015}).

\bibitem{2016:oe:tanida}
\bibinfo{author}{Horisaki, R.}, \bibinfo{author}{Takagi, R.} \&
  \bibinfo{author}{Tanida, J.}
\newblock \bibinfo{title}{Learning-based imaging through scattering media}.
\newblock \emph{\bibinfo{journal}{Optics Express}}
  \textbf{\bibinfo{volume}{24}}, \bibinfo{pages}{13738--13743}
  (\bibinfo{year}{2016}).

\bibitem{2017:ao:tanida}
\bibinfo{author}{Horisaki, R.}, \bibinfo{author}{Takagi, R.} \&
  \bibinfo{author}{Tanida, J.}
\newblock \bibinfo{title}{Learning-based focusing through scattering media}.
\newblock \emph{\bibinfo{journal}{Applied Optics}}
  \textbf{\bibinfo{volume}{56}}, \bibinfo{pages}{4358--4362}
  (\bibinfo{year}{2017}).

\bibitem{2017:arxiv:barbastathis}
\bibinfo{author}{Li, S.}, \bibinfo{author}{Deng, M.}, \bibinfo{author}{Lee,
  J.}, \bibinfo{author}{Sinha, A.} \& \bibinfo{author}{Barbastathis, G.}
\newblock \bibinfo{title}{Imaging through glass diffusers using densely
  connected convolutional networks}.
\newblock \emph{\bibinfo{journal}{arXiv preprint}}
  \textbf{\bibinfo{volume}{arXiv:1711.06810v1}} (\bibinfo{year}{2017}).

\bibitem{2017:arxiv:lyu}
\bibinfo{author}{Lyu, M.}, \bibinfo{author}{Wang, H.}, \bibinfo{author}{Li, G.}
  \& \bibinfo{author}{Situ, G.}
\newblock \bibinfo{title}{Exploit imaging through opaque wall via deep
  learning}.
\newblock \emph{\bibinfo{journal}{arXiv preprint}}
  \textbf{\bibinfo{volume}{arXiv.1711.06810v1}} (\bibinfo{year}{2017}).

\bibitem{1990:nature:sandler}
\bibinfo{author}{\color{\changeshighlight}J. R.~P.~Angel},
  \bibinfo{author}{Wizinowich, P.}, \bibinfo{author}{Lloyd-Hart, M.} \&
  \bibinfo{author}{Sandler, D.}
\newblock \bibinfo{title}{Adaptive optics for array telescopes using
  neural-network techniques}.
\newblock \emph{\bibinfo{journal}{Nature}} \textbf{\bibinfo{volume}{348}},
  \bibinfo{pages}{221--224} (\bibinfo{year}{1990 \color{black}}).

\bibitem{2018:prl:carminati}
\bibinfo{author}{\color{\changeshighlight} N.~Fayard},
  \bibinfo{author}{Goetschy, A.}, \bibinfo{author}{Pierrat, R.} \&
  \bibinfo{author}{Carminati, R.}
\newblock \bibinfo{title}{Mutual information between reflected and transmitted
  speckle images}.
\newblock \emph{\bibinfo{journal}{Physical Review Letters}}
  \textbf{\bibinfo{volume}{120}}, \bibinfo{pages}{073901} (\bibinfo{year}{2018
  \color{black}}).

\bibitem{2015:mc:betzig}
\bibinfo{author}{\color{\changeshighlight} Zhe~Liu}, \bibinfo{author}{Lavis,
  L.~D.} \& \bibinfo{author}{Betzig, E.}
\newblock \bibinfo{title}{Imaging live-cell dynamics and structure at the
  single-molecule level}.
\newblock \emph{\bibinfo{journal}{Molecular Cell}}
  \textbf{\bibinfo{volume}{58}}, \bibinfo{pages}{644--659} (\bibinfo{year}{2015
  \color{black}}).

\bibitem{2004:ol:denk}
\bibinfo{author}{\color{\changeshighlight} M.~Feierabend},
  \bibinfo{author}{Ruckel, M.} \& \bibinfo{author}{Denk, W.}
\newblock \bibinfo{title}{Coherence-gated wave-front sensing in strongly
  scattering samples}.
\newblock \emph{\bibinfo{journal}{Optics Letters}}
  \textbf{\bibinfo{volume}{29}}, \bibinfo{pages}{2255--2257}
  (\bibinfo{year}{2004 \color{black}}).

\bibitem{2012:oe:cui}
\bibinfo{author}{\color{\changeshighlight} R.~Fiolka}, \bibinfo{author}{Si, K.}
  \& \bibinfo{author}{Cui, M.}
\newblock \bibinfo{title}{Parallel wavefront measurements in ultrasound pulse
  guided digital phase conjugation}.
\newblock \emph{\bibinfo{journal}{Optics Express}}
  \textbf{\bibinfo{volume}{20}}, \bibinfo{pages}{24827--24834}
  (\bibinfo{year}{2012 \color{black}}).

\bibitem{2015:book:drexler}
\bibinfo{author}{Drexler, W.} \& \bibinfo{author}{Fujimoto, J.~G.}
\newblock \emph{\bibinfo{title}{\color{\changeshighlight} Optical Coherence
  Tomography}} (\bibinfo{publisher}{Springer International Publishing},
  \bibinfo{year}{2015 \color{black}}).

\bibitem{2015:np:choi}
\bibinfo{author}{\color{\changeshighlight} S.~Kang} \emph{et~al.}
\newblock \bibinfo{title}{Imaging deep within a scattering medium using
  collective accumulation of single-scattered waves}.
\newblock \emph{\bibinfo{journal}{Nature Photonics}}
  \textbf{\bibinfo{volume}{9}}, \bibinfo{pages}{253--258} (\bibinfo{year}{2015
  \color{black}}).

\bibitem{2016:sa:aubry}
\bibinfo{author}{\color{\changeshighlight} A.~Badon} \emph{et~al.}
\newblock \bibinfo{title}{Smart optical coherence tomography for ultra-deep
  imaging through highly scattering media}.
\newblock \emph{\bibinfo{journal}{Science Advances}}
  \textbf{\bibinfo{volume}{2}}, \bibinfo{pages}{e1600370} (\bibinfo{year}{2016
  \color{black}}).

\bibitem{2017:nc:choi}
\bibinfo{author}{\color{\changeshighlight} S.~Kang} \emph{et~al.}
\newblock \bibinfo{title}{High-resolution adaptive optical imaging within thick
  scattering media using closed-loop accumulation of single scattering}.
\newblock \emph{\bibinfo{journal}{Nature Communications}}
  \textbf{\bibinfo{volume}{8}}, \bibinfo{pages}{2157} (\bibinfo{year}{2017
  \color{black}}).

\bibitem{2018:optica:judkewitz}
\bibinfo{author}{\color{\changeshighlight} M.~Kadobianskyi},
  \bibinfo{author}{Papadopoulos, I.~N.}, \bibinfo{author}{Chaigne, T.},
  \bibinfo{author}{Horstmeyer, R.} \& \bibinfo{author}{Judkewitz, B.}
\newblock \bibinfo{title}{Scattering correlations of time-gated light}.
\newblock \emph{\bibinfo{journal}{Optica}} \textbf{\bibinfo{volume}{5}},
  \bibinfo{pages}{389--394} (\bibinfo{year}{2018 \color{black}}).

\bibitem{2015:pra:carminati}
\bibinfo{author}{\color{\changeshighlight} N.~Fayard}, \bibinfo{author}{Caze,
  A.}, \bibinfo{author}{Pierrat, R.} \& \bibinfo{author}{Carminati, R.}
\newblock \bibinfo{title}{Intensity correlations between reflected and
  transmitted speckle patterns}.
\newblock \emph{\bibinfo{journal}{Physical Review A}}
  \textbf{\bibinfo{volume}{92}}, \bibinfo{pages}{033827} (\bibinfo{year}{2015
  \color{black}}).

\bibitem{2017:arxiv:bertolotti}
\bibinfo{author}{\color{\changeshighlight} I.~Starshynov} \emph{et~al.}
\newblock \bibinfo{title}{Non-gaussian correlations between reflected and
  transmitted intensity patterns emerging from opaque disordered media}.
\newblock \emph{\bibinfo{journal}{Physical Review X}}
  \textbf{\bibinfo{volume}{8}}, \bibinfo{pages}{021041} (\bibinfo{year}{2018
  \color{black}}).

\bibitem{2013:prl:Yaqoob}
\bibinfo{author}{\color{\changeshighlight} Youngwoon~Choi} \emph{et~al.}
\newblock \bibinfo{title}{Measurement of the time-resolved reflection matrix
  for enhancing light energy delivery into a scattering medium}.
\newblock \emph{\bibinfo{journal}{Physical Review Letters}}
  \textbf{\bibinfo{volume}{111}}, \bibinfo{pages}{243901} (\bibinfo{year}{2013
  \color{black}}).

\bibitem{2015:oc:park}
\bibinfo{author}{\color{\changeshighlight} Hyeonseung~Yu},
  \bibinfo{author}{Park, J.~H.} \& \bibinfo{author}{Park, Y.}
\newblock \bibinfo{title}{Measuring large optical reflection matrices of turbid
  media}.
\newblock \emph{\bibinfo{journal}{Optics Communications}}
  \textbf{\bibinfo{volume}{352}}, \bibinfo{pages}{33–38} (\bibinfo{year}{2015
  \color{black}}).

\bibitem{2018:np:Choi}
\bibinfo{author}{\color{\changeshighlight} Seungwon~Jeong} \emph{et~al.}
\newblock \bibinfo{title}{Focusing of light energy inside a scattering medium
  by controlling the time-gated multiple light scattering}.
\newblock \emph{\bibinfo{journal}{Nature Photonics}}
  \textbf{\bibinfo{volume}{12}}, \bibinfo{pages}{277–283}
  (\bibinfo{year}{2018 \color{black}}).

\bibitem{2013:josa:rand}
\bibinfo{author}{\color{\changeshighlight} Curtis~Jin},
  \bibinfo{author}{Nadakuditi, R.~R.}, \bibinfo{author}{Michielssen, E.} \&
  \bibinfo{author}{Rand, S.~C.}
\newblock \bibinfo{title}{Iterative, backscatter-analysis algorithms for
  increasing transmission and focusing light through highly scattering random
  media}.
\newblock \emph{\bibinfo{journal}{J. Opt. Soc. Am. A}}
  \textbf{\bibinfo{volume}{30}}, \bibinfo{pages}{1592--1602}
  (\bibinfo{year}{2013 \color{black}}).

\bibitem{2014:josa:rand}
\bibinfo{author}{\color{\changeshighlight} Curtis~Jin},
  \bibinfo{author}{Nadakuditi, R.~R.}, \bibinfo{author}{Michielssen, E.} \&
  \bibinfo{author}{Rand, S.~C.}
\newblock \bibinfo{title}{Backscatter analysis based algorithms for increasing
  transmission through highly scattering random media using
  phase-only-modulated wavefronts}.
\newblock \emph{\bibinfo{journal}{J. Opt. Soc. Am. A}}
  \textbf{\bibinfo{volume}{31}}, \bibinfo{pages}{1788--1800}
  (\bibinfo{year}{2014 \color{black}}).

\bibitem{2011:oe:mosk}
\bibinfo{author}{Akbulut, D.}, \bibinfo{author}{Huisman, T.~J.},
  \bibinfo{author}{van Putten, E.~G.}, \bibinfo{author}{Vos, W.~L.} \&
  \bibinfo{author}{Mosk, A.~P.}
\newblock \bibinfo{title}{Focusing light through random photonic media by
  binary amplitude modulation}.
\newblock \emph{\bibinfo{journal}{Optics Express}}
  \textbf{\bibinfo{volume}{19}}, \bibinfo{pages}{4017--4029}
  (\bibinfo{year}{2011}).

\bibitem{2014:nn:schmidhuber}
\bibinfo{author}{\color{\changeshighlight} Jurgen~Schmidhuber}.
\newblock \bibinfo{title}{Deep learning in neural networks: An overview}.
\newblock \emph{\bibinfo{journal}{Neural Networks}}
  \textbf{\bibinfo{volume}{61}}, \bibinfo{pages}{85--117} (\bibinfo{year}{2014
  \color{black}}).

\bibitem{2009:njp:boshier}
\bibinfo{author}{Henderson, K.}, \bibinfo{author}{Ryu, C.},
  \bibinfo{author}{MacCormick, C.} \& \bibinfo{author}{Boshier, M.~G.}
\newblock \bibinfo{title}{Experimental demonstration of painting arbitrary and
  dynamic potentials for bose–einstein condensates}.
\newblock \emph{\bibinfo{journal}{New Journal of Physics}}
  \textbf{\bibinfo{volume}{11}}, \bibinfo{pages}{043030}
  (\bibinfo{year}{2009}).

\bibitem{2009:pnas:tank}
\bibinfo{author}{\color{\changeshighlight} John Peter~Rickgauer} \&
  \bibinfo{author}{Tank, D.~W.}
\newblock \bibinfo{title}{Two-photon excitation of channelrhodopsin-2 at
  saturation}.
\newblock \emph{\bibinfo{journal}{PNAS}} \textbf{\bibinfo{volume}{106}},
  \bibinfo{pages}{15025--15030} (\bibinfo{year}{2009 \color{black}}).

\bibitem{2017:oe:piestun}
\bibinfo{author}{Caravaca-Aguirre, A.~M.} \& \bibinfo{author}{Piestun, R.}
\newblock \bibinfo{title}{Single multimode fiber endoscope}.
\newblock \emph{\bibinfo{journal}{Optics Express}}
  \textbf{\bibinfo{volume}{25}}, \bibinfo{pages}{1656--1665}
  (\bibinfo{year}{2017}).

\bibitem{2017:arxiv:piestun}
\bibinfo{author}{Ohayon, S.}, \bibinfo{author}{Caravaca-Aguirre, A.~M.},
  \bibinfo{author}{Piestun, R.} \& \bibinfo{author}{DiCarlo, J.~J.}
\newblock \bibinfo{title}{Deep brain fluorescence imaging with minimally
  invasive ultra-thin optical fibers}.
\newblock \emph{\bibinfo{journal}{arxiv preprint}}
  \textbf{\bibinfo{volume}{arXiv:1703.07633}} (\bibinfo{year}{2017}).

\bibitem{2018:oe:Zhao}
\bibinfo{author}{Zhao, T.}, \bibinfo{author}{Deng, L.}, \bibinfo{author}{Wang,
  W.}, \bibinfo{author}{Elson, D.~S.} \& \bibinfo{author}{Su, L.}
\newblock \bibinfo{title}{Bayes’ theorem-based binary algorithm for fast
  reference-less calibration of a multimode fiber}.
\newblock \emph{\bibinfo{journal}{Optics Express}}
  \textbf{\bibinfo{volume}{26}}, \bibinfo{pages}{20368--20378}
  (\bibinfo{year}{2018}).

\bibitem{2012:book:goshtasby}
\bibinfo{author}{Goshtasby, A.~A.}
\newblock \emph{\bibinfo{title}{\color{\changeshighlight} Image Registration}}
  (\bibinfo{publisher}{Springer}, \bibinfo{year}{2012 \color{black}}).

\bibitem{2015:spie:loterie}
\bibinfo{author}{Damien~Loterie, D. P. C.~M., Salma~Farahi}.
\newblock \bibinfo{title}{Complex pattern projection through a multimode fiber}
  (\bibinfo{year}{2015}).

\bibitem{2018:lsa:booth}
\bibinfo{author}{Sun, B.} \emph{et~al.}
\newblock \bibinfo{title}{Four-dimensional light shaping: manipulating
  ultrafast spatiotemporal foci in space and time}.
\newblock \emph{\bibinfo{journal}{Light: Science and Applications}}
  \textbf{\bibinfo{volume}{7}}, \bibinfo{pages}{17117} (\bibinfo{year}{2018}).

\bibitem{2017:optica:silberberg}
\bibinfo{author}{Frostig, H.} \emph{et~al.}
\newblock \bibinfo{title}{Focusing light by wavefront shaping through disorder
  and nonlinearity}.
\newblock \emph{\bibinfo{journal}{Optica}} \textbf{\bibinfo{volume}{4}},
  \bibinfo{pages}{1073--1079} (\bibinfo{year}{2017}).

\bibitem{2011:np:wang}
\bibinfo{author}{\color{\changeshighlight} X.~A~Xu}, \bibinfo{author}{Liu,
  H.~L.} \& \bibinfo{author}{Wang, L.~V.}
\newblock \bibinfo{title}{Time-reversed ultrasonically encoded optical focusing
  into scattering media}.
\newblock \emph{\bibinfo{journal}{Nature Photonics}}
  \textbf{\bibinfo{volume}{5}}, \bibinfo{pages}{154--157} (\bibinfo{year}{2011
  \color{black}}).

\bibitem{2018:ol:Kuschmierz}
\bibinfo{author}{Kuschmierz, R.}, \bibinfo{author}{Scharf, E.},
  \bibinfo{author}{Koukourakis, N.} \& \bibinfo{author}{Czarske, J.~W.}
\newblock \bibinfo{title}{Self-calibration of lensless holographic endoscope
  using programmable guide stars}.
\newblock \emph{\bibinfo{journal}{Optics Letters}}
  \textbf{\bibinfo{volume}{43}}, \bibinfo{pages}{2997--3000}
  (\bibinfo{year}{2018}).

\bibitem{2018:arxiv:tian}
\bibinfo{author}{Li, L.~T.~Y.}, \bibinfo{author}{Xue, Y.} \&
  \bibinfo{author}{Tian, L.}
\newblock \bibinfo{title}{Deep speckle correlation: a deep learning approach
  towards scalable imaging through scattering media}.
\newblock \emph{\bibinfo{journal}{Optica}} \textbf{\bibinfo{volume}{5}},
  \bibinfo{pages}{1181--1190} (\bibinfo{year}{2018}).

\bibitem{chollet2015keras}
\bibinfo{author}{Chollet, F.} \emph{et~al.}
\newblock \bibinfo{title}{Keras}.
\newblock \bibinfo{howpublished}{\url{https://github.com/keras-team/keras}}
  (\bibinfo{year}{2015}).

\bibitem{tensorflow2015-whitepaper}
\bibinfo{author}{Abadi, M.} \emph{et~al.}
\newblock \bibinfo{title}{{TensorFlow}: Large-scale machine learning on
  heterogeneous systems} (\bibinfo{year}{2015}).
\newblock \urlprefix\url{http://tensorflow.org/}.

\end{thebibliography}



\end{document}